\documentclass[12pt,preprint]{aastex}

\usepackage{graphics,epsfig}

\shorttitle{Multiwavelength Study of Faint X-ray Sources in the SWIRE Survey}
\shortauthors{Franceschini et al.}

\received{2004 June 15}
\begin{document}

\title{A Complete Multiwavelength Characterization of Faint Chandra X-ray Sources Seen in the Spitzer Wide-Area IR Extragalactic (SWIRE) Survey}

\author{Alberto Franceschini, James Manners}
\affil{Dipartimento di Astronomia, Universita di Padova, Vicolo 
Osservatorio 5, I-35122 Padua, Italy} 

\author{Maria del Carmen Polletta}
\affil{Center for Astrophysics \& Space Sciences, University of 
California, San Diego, La Jolla, CA 92093--0424, USA}

\author{Carol Lonsdale}
\affil{Infrared Processing \& Analysis Center, California Institute 
of Technology, 100-22, Pasadena, CA 91125, USA} 

\author{Eduardo Gonzalez-Solares }
\affil{Cambridge, University of Cambridge, UK}

\author{Jason Surace, Dave Shupe, Fan Fang, C. Kevin Xu, \& Duncan Farrah}
\affil{Infrared Processing \& Analysis Center, California Institute 
of Technology, 100-22, Pasadena, CA 91125, USA}

\author{Stefano Berta \& Giulia Rodighiero}
\affil{Dipartimento di Astronomia, Universita di Padova, Vicolo 
Osservatorio 5, I-35122 Padua, Italy}

\author{Ismael Perez-Fournon \& Evanthia Hatziminaoglou}
\affil{Instituto de Astrofisica de Canarias, 38200 La Laguna, 
Tenerife, Spain} 

\author{Harding E. Smith \& Brian Siana}
\affil{Center for Astrophysics \& Space Sciences, University of 
California, San Diego, La Jolla, CA 92093--0424, USA} 

\author{Michael Rowan-Robinson, Kirpal Nandra, Tom Babbedge \& Mattia Vaccari}
\affil{Astrophysics Group, Blackett Laboratory, Imperial College,
Prince Consort Road, London, SW7 2BW, UK} 

\author{Seb Oliver}
\affil{Astronomy Centre, University of Sussex, Brighton BN1 9QJ, UK}

\author{Belinda Wilkes }
\affil{Harvard-Smithsonian Center for Astrophysics, Cambridge, USA} 

\author{Frazer Owen}
\affil{National Radio Astronomy Observatory, P.O. Box O, Socorro, 
NM 87801, USA} 

\author{Deborah Padgett, Dave Frayer, Tom Jarrett, Frank Masci}
\affil{Infrared Processing \& Analysis Center, California Institute 
of Technology, 100-22, Pasadena, CA 91125, USA}



\author{Gordon Stacey}
\affil{Department of Astronomy, Cornell University, 220 Space 
Science Building, Ithaca, NY 14853, USA} 




\author{Omar Almaini}
\affil{School of Physics and Astronomy, University of Nottingham, University
Park Nottingham NG7 2RD, UK
}

\author{Richard McMahon}
\affil{Institute of Astronomy, Madingley Road, Cambridge CB3 0HA, UK
}

\author{Olivia Johnson, Andrew Lawrence}
\affil{Institute for Astronomy, University of Edinburgh, Royal Observatory,
Blackford Hill, Edinburgh EH9 3HJ, UK
}

\author{Chris Willott}
\affil{Herzberg Institute of Astrophysics, National Research Council, 5071 West
Saanich Rd, Victoria, B.C. V9E 2E7, Canada
}



\begin{abstract}

We exploit deep combined observations with Spitzer and Chandra of the Spitzer Wide-area Infrared Extragalactic Survey (SWIRE) in the ELAIS-N1
region, to investigate the nature of the faint X-ray and IR sources in common, to identify AGN/starburst diagnostics, and to study the sources of the X-ray and IR cosmic backgrounds (XRB and CIRB). In the 17'$\times$17' area of the Chandra ACIS-I image there are approximately 3400 SWIRE near-IR sources with 4 sigma detections in at least 2 IRAC bands and 988 sources detected at 24$\mu$m with the Multiband Imaging Photometer (MIPS) brighter than $S_{24}\simeq 0.1\ mJy$. Of these, 102 IRAC and 59 MIPS sources have Chandra counterparts, out of a total of 122 X-ray sources present in the area with $S_{0.5-8 kev}>10^{-15}\ erg/cm^2/s$. We have constructed Spectral Energy Distributions (SEDs) for each source using data from the 4 IRAC wavebands, Chandra fluxes in the hard (2 - 8 keV) and soft (0.5 - 2 keV) X-rays, and optical follow-up data in the wavebands U, g', r', i', Z, and H. 
We fit a number of spectral templates to the SEDs at optical and infrared wavelengths to determine photometric redshifts and spectral categories, and also make use of diagnostics based on the X-ray luminosities, hardness ratios, X-ray to infrared spectral slopes and optical morphologies. 
Although we have spectroscopic redshifts for only a minority of the Chandra sources (10 type-1 QSOs or Seyferts and 3 galaxies), the available SEDs  constrain the redshifts for most of the sample sources, which turn out to be typically at $0.5<z<2$. 
We find that 39\% of the Chandra sources are dominated by type-1 AGN emission (QSOs or Seyfert 1), 23\% display optical/IR spectra typical of type-2 AGNs, while the remaining 38\% fraction show starburst-like or even normal galaxy spectra (including 5 passively evolving early-type galaxies). Since we prove that all these galaxies are dominated by AGN emission in X-rays (considering their large 0.5-8 keV rest-frame X-ray luminosities
and their high X-ray to IR flux ratios), this brings the fraction of type-1 AGNs to be
80\% of the type-2: even assuming that all the Chandra sources undetected
by Spitzer are type-2 AGNs, the type-1 fraction would exceed 1/3 of the total population.
Our analysis of the mid-IR MIPS 24 $\mu$m selected sources, composing $\sim 50\%$ of the CIRB background, shows that the fraction of those dominated by an AGN (either type-1 or type-2) is relatively constant with the IR flux and around 10-15\%. Our combined IR and hard-X observations allow us to verify that the dust covering fraction in type-1 AGNs is widely distributed between $\sim$10 and 100\%.
A significant fraction, from 15 to 30\%, or more, of the sources of the XRB are hosted in galaxies whose optical-IR spectra are dominated by starburst (or normal galaxy) emission and for which only the hard X-ray spectra reveal the presence of a moderately luminous hidden AGN. 

\end{abstract}


\section{Introduction}

The Spitzer Space Telescope (SST) is providing a new sensitive tool for cosmological investigations over a wide spectral region from 3 to 160 $\mu$m. The Spitzer
Wide-area InfraRed Extragalactic Legacy Survey, SWIRE, in particular, with its large areal and IR spectral coverage, will not only allow us to quantify the role of the environment in structure formation and evolution, but is also providing rich enough statistical samples for investigating relatively rare source populations (Lonsdale et al. 2003, 2004).

The sources of the cosmic X-ray background, efficiently detected by Chandra and XMM-Newton, belong to this last category of relatively uncommon objects. Recently, the importance of combined hard X-ray and mid- and far-IR data to test the AGN unification model and to verify the standard obscured accretion paradigm for the origin of the XRB has been emphasized
(Risaliti et al. 2000; Franceschini et al. 2002; Fadda et al. 2002; Alonso-Herrero et al. 2004; Rigby et al 2004; Manners et al. 2004; Alexander et al. 2003).
Indeed, a relationship has been suggested to hold between the XRB and the more recently discovered cosmological background in the far-IR, the CIRB (Puget et al. 1996; Hauser et al. 1998).
The X-ray emission that is photoelectrically absorbed in type-2 AGNs dominating the XRB is expected to be downgraded in energy by the dusty circumnuclear
medium and to emerge thermally reprocessed in the IR between a few and a few hundred microns (e.g. Granato, Danese, \& Franceschini 1997). 
An important test of this picture is then provided by detection in the IR of the downgraded energy absorbed in X-rays.

A related aspect is to investigate how the X-ray and UV-optical absorbed flux from AGNs contributes to explaining the energetics of the IR sources and of the whole IR background itself.
Considering that the CIRB contains more than half of the total radiant energy produced by cosmic sources (Hauser \& Dwek 1998; Franceschini et al. 2001), it is of crucial importance to assess the physical origin of this component. 
The issue is still relatively unsettled, the expected AGN fraction ranging from 50\% of the CIRB due to AGNs (under some extreme assumptions about the IR emissivity of type-2 objects and the percentage of Compton-thick sources; see Almaini, Lawrence, \& Boyle 1999; Fabian \& Iwasawa 1999), to a few percent (Elbaz et al. 2002).

First attempts to answer these questions were based on the pioneering mid-IR sky surveys with the ISO observatory. Fadda et al. (2002) detected 24 sources in common between ISO at 15 $\mu$m and XMM-Newton in an area of the Lockman Hole. Based on that, Fadda et al. and Franceschini et al. (2002) concluded that the expectations of the obscured accretion model are basically met. These studies also suggested that only a minor fraction of the CIRB seems due to gravitational accretion. 
These conclusions were however limited by small-number statistics and by poor spectral coverage of the source SEDs (essentially limited to the 15 $\mu$m datapoint).

Also deep long-wavelength surveys with SCUBA and MAMBO have been cross-correlated with ultra-deep Chandra images, with the result that, although many high-z sub-mm SCUBA sources show detectable hard X-ray emission, the large majority of the bolometric flux does not appear to be due to an AGN (Alexander et al. 2004). In any case not much can be concluded from these data about the origin of the CIRB because it is still uncertain how far the sub-mm sources contribute to the CIRB at its peak at around 100$\mu$m. This is also made uncertain by a quite complicated source selection function (including sub-mm and radio/optical identification), and again poor statistics.

We report in this paper a refined analysis of the relationship between X-ray and IR emission for faint cosmic sources over a wide redshift interval based on combined Spitzer and Chandra observations of an area in ELAIS-N1. This region has been previously observed with ISO as part of the ELAIS survey (Oliver et al. 2000). ISO detected 6 extragalactic sources at 15 $\mu$m within the 17'x17' area of the ELAIS-N1 Chandra ACIS-I image, three of which have detectable counterparts in X-rays according to Manners et al. (2004).

For comparison, the present SWIRE survey in the same small area detects $\sim$900 sources at 24 $\mu$m above $S_{24}=100\ \mu Jy$ and 3500 IRAC near-IR galaxies. We find that 102 (84\%) of the 122 Chandra sources in the ACIS-I area now have mid and far IR counterparts from Spitzer.

The Spitzer Space Telescope, Chandra and ancillary optical observations of the ELAIS-N1 region are detailed in Sect. \ref{sec:observations}. Particular attention is paid here to the estimation of photometric redshifts for our sample sources, given that only a fraction of them have optical spectroscopic data. Although our approach is similar to others recently published for X-ray selected source populations (Franceschini et al. 2002; Zheng et al. 2004), our advantage here is to benefit from an accurate coverage of the near- and mid-IR spectrum with the 4 IRAC channels centered at 3.6, 4.5, 5.8, 8 $\mu$m and the MIPS 24 $\mu$m channel.

The main results of our analysis are reported in Section \ref{sec:results}, in which we detail the X-ray and IR cross-correlation and source identification procedures, the analysis of the SEDs and classification of the sources, the X-ray to IR flux correlations and color and hardness-ratio plots. We discuss in particular new diagnostic diagrams for AGN- and galaxy-dominated emission in the optical-IR based on structural differences in the spectra. Photometric redshifts are used to infer important statistical information on redshift and luminosity distributions. The X-ray and IR identification statistics and some inferences about the AGN dust covering, as well as a comparison of Chandra-detected and undetected SWIRE populations, are also addressed in Section \ref{sec:results}. The paper's conclusions are summarized in Section \ref{con}.

\section{OBSERVATIONS}\label{sec:observations}

\subsection{Spitzer Space Telescope Observations}

The SWIRE ELAIS-N1 field was imaged by the IRAC multiband camera on Spitzer in January 2004 and with MIPS in early February 2004, following the strategy described in Lonsdale et al. (2004). The original observing plan was corrected after the Spitzer PV phase, by increasing the depth of the MIPS coverage to account for a reduced efficiency of the 70 $\mu$m channel. 
The observations were centered at the position $16^h00^m, +59^d01^m$.

Data processing began with the Spitzer Basic Calibrated Data (BCD), corrected for bias offsets and pixel-to-pixel gain variations (flat-fielding), and 
flux-calibrated in surface brightness.  Additional IRAC processing corrected for latent images and electronic offsets.  For MIPS, scan-mirror-dependent flats were derived from the data and applied to the BCD images.  Individual images, which have measurable spatial distortions, were reprojected onto a single common projection system on the sky, and coadded after correction for cosmic rays and other transient artifacts.  

Fluxes were extracted in 5.8\arcsec\ apertures for IRAC ($\sim$2-3$\times$ the FWHM beam) and 12\arcsec\ for MIPS 24$\mu$m, using SExtractor.  
Sources are typically unresolved by the large Spitzer beams ($>$2\arcsec\ at the shortest wavelength).    The absolute flux calibrations are believed to be correct within roughly 10\% for the IRAC and for the MIPS 24$\mu$m channel data and were confirmed for IRAC by comparison to 2MASS.
The resulting catalogs were examined by eye and remaining spurious sources (radiation, scattered light, etc.) were removed by hand.
More details on the data processing are given in Surace et al (2004) and Shupe et al. (2004).

\begin{figure}
 \includegraphics[width=\textwidth,height=1.1\textwidth]{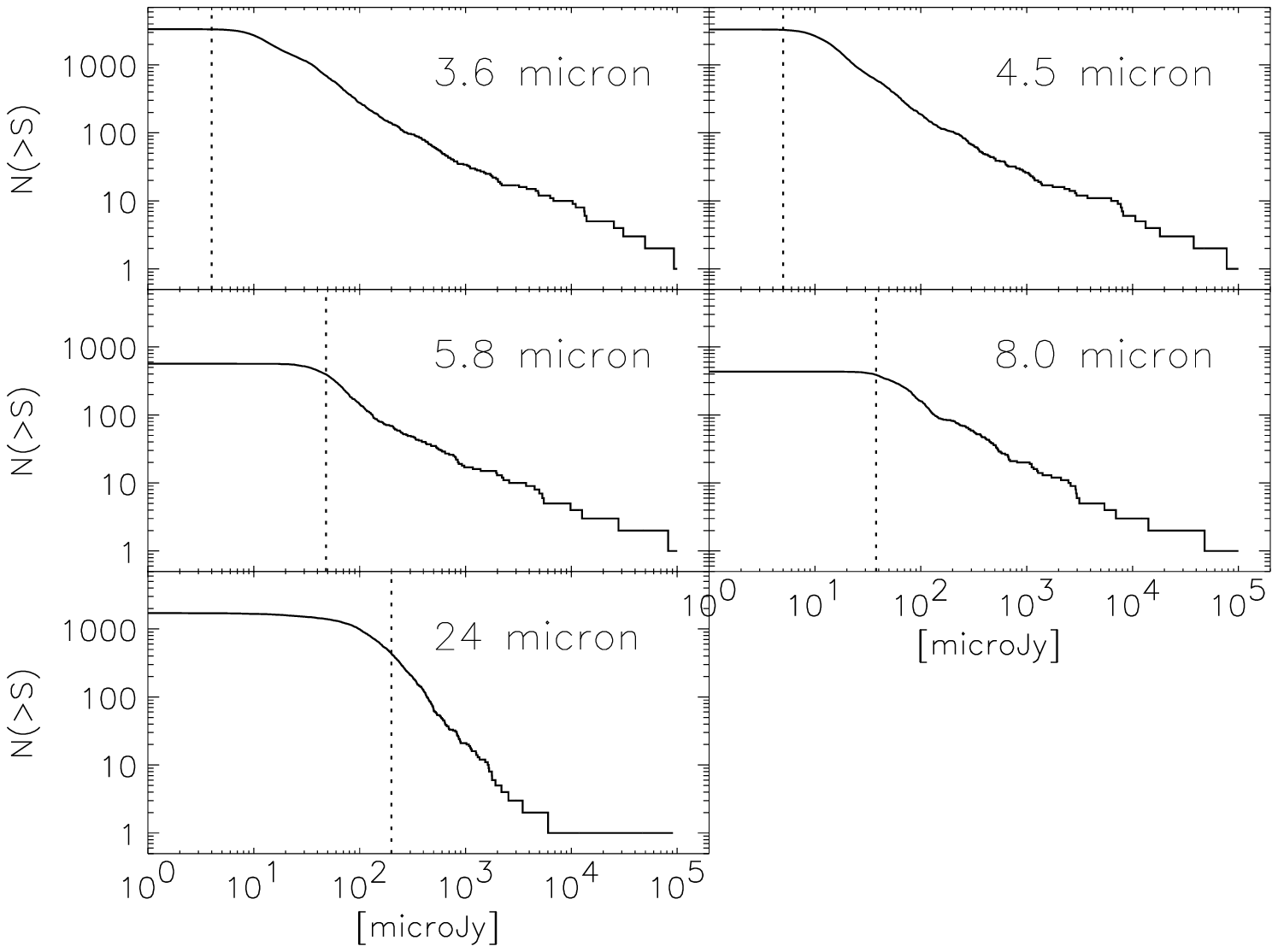}
\caption{Cumulative number counts of galaxies detected with $S/N>4$ by Spitzer within the 286 sq. arcmin of the Chandra ACIS-I image.  The vertical dotted lines mark the limiting fluxes for 90\% completeness.
}
\label{fig1}
\end{figure}

   
The achieved sensitivities with the SWIRE survey in ELAIS-N1 turned out to be deeper compared with pre-flight expectations in the two short-wavelength IRAC channels and in the MIPS 24 $\mu$m one, while performing less well in the other bands.
The integral number counts of Spitzer sources in the Chandra ACIS-I 286 sq. arcmin field are reported in Fig. \ref{fig1}. Dotted vertical lines mark the flux limits corresponding
to 90\% completeness limits as determined from the deviation of the observed number counts from a power law and from simulations (Shupe et al. 2004).



\subsection{The X-ray Data}\label{X}

The X-ray observations were taken as part of the ELAIS Deep X-ray Survey
(EDXS), and are described in detail in Manners {\it et  al.} (2003). For
this analysis we use the {\it Chandra} Advanced CCD Imaging
Spectrometer (ACIS) observation of $75$ks centred on 16:10:20.11 +54:33:22.3 (J2000) in the ELAIS-N1 region.
The aimpoint was focused on the ACIS-I chips which consist of 4 CCDs arranged in a $2 \times 2$ array covering an area of $16.9' \times 16.9'$ (286 sq. arcmin.). Bad pixels and columns were removed and data were filtered to eliminate high
background times (due to strong solar flares), leaving 71.5 ks of good data after filtering. Source
detection and characterisation in 3 bands (0.5 - 8 keV [full band], 0.5 - 2 keV [soft band], and 2 - 8 keV [hard band]) were achieved using the {\small
WAVDETECT} software package (Freeman {\it et al.} 2002). For each band, exposure maps were constructed to account for variations in effective exposure across an image, and used to remove bias from the source detection
and to calculate source fluxes. Counts to photon calibration assumed a standard power-law model spectrum, with photon index $\Gamma = 1.7$.

Sources were detected to flux levels of $2.3 \times 10^{-15}$
erg s$^{-1}$ cm$^{-2}$ in the 0.5 - 8 keV band, $9.4 \times 10^{-16}$
erg s$^{-1}$ cm$^{-2}$ in the 0.5 - 2 keV band, and $5.2 \times
10^{-15}$ erg s$^{-1}$ cm$^{-2}$ in the 2 - 8 keV band. Sources are detectable to these flux limits over 90\% of the nominal survey area. For this
analysis we used sources detected in the full band of ACIS-I only, which are 122 in the N1 region.  Of the 102 sources in common between Chandra and SWIRE, 83 have significant detections in the separate soft X-ray band (0.5 - 2 keV), and 64 are detected in the hard (2 - 8 keV) band.

\begin{figure}
 \includegraphics[width=\textwidth]{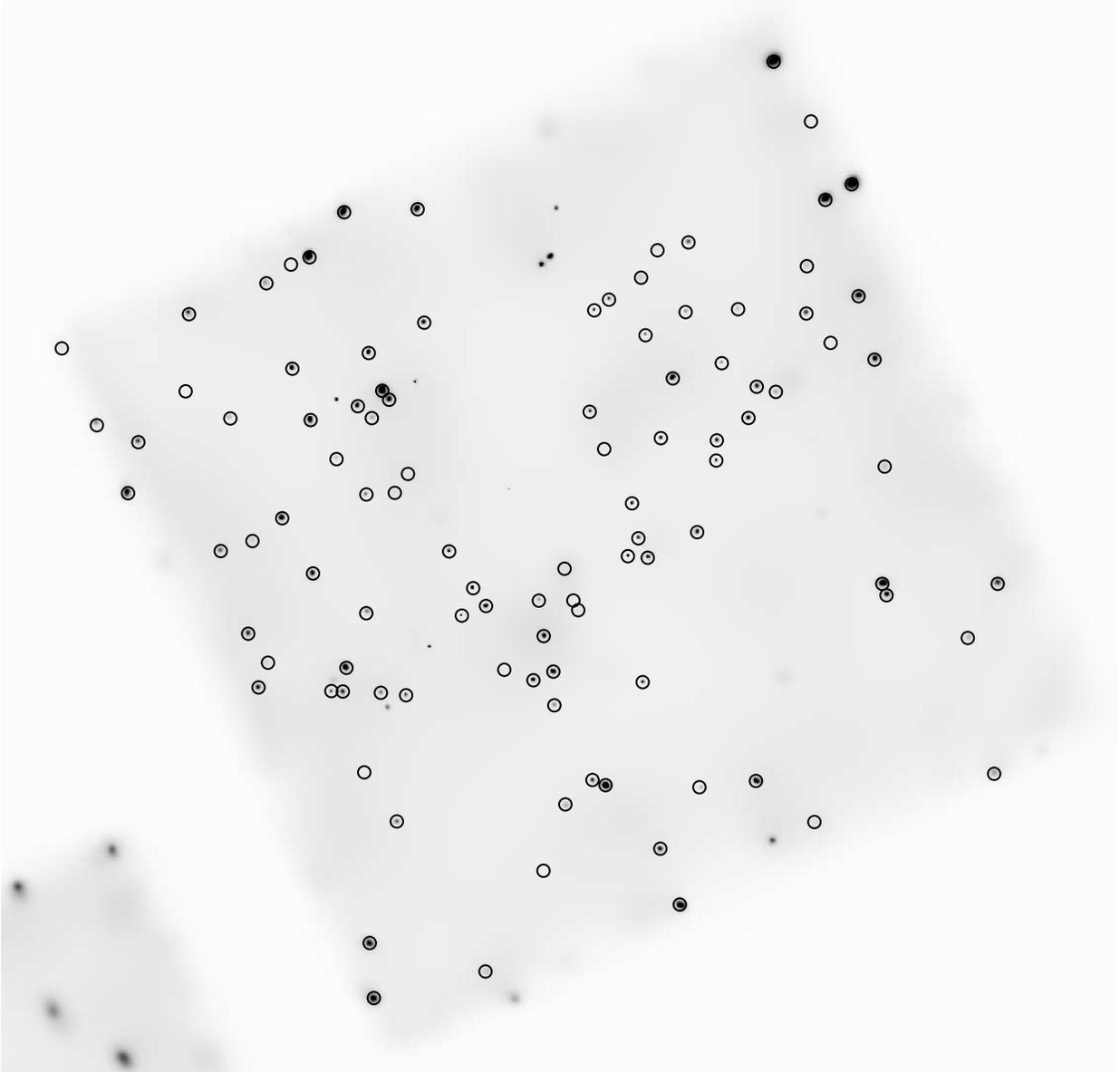}
\caption{The adaptively smoothed Chandra image overlayed with positions for 102 sources detected by Spitzer with the IR Array Camera (IRAC).
}
\label{fig2}
\end{figure}

The smoothed 0.5 - 8 keV Chandra ACIS-I image is reported in Fig. \ref{fig2} with the associated SWIRE sources overlayed. The image has been adaptively smoothed using the flux-conserving algorithm CSMOOTH from the 'Chandra Interactive Analysis of Observations' (CIAO) package.
We defer to Manners {\it et  al.} (2003 and 2004) for more details on the X-ray data reduction and analysis.

\subsection{The Optical Data}

Follow-up observations of the EDXS sources have been taken in the
bands U, g, r, i, and H and are detailed in Gonzalez-Solares {\it et al.} (2004). The Wide Field Camera (WFC) on the Isaac Newton Telescope (INT)
was used to observe a $30' \times 30'$ region centred on the Chandra
N1 pointing. Data in the g, r, and i bands were observed in July 2000 and the
U band in June 2001 using the Sloan Digital Sky Survey (SDSS)
photometric system. Total integration times were 100 minutes in the g,
r, and i bands and 120 minutes in the U band, reaching Vega magnitude
limits for a point like source in an aperture of $3''$ of $U=24.0$,
$g=25.5$, $r=25.2$, $i=24.1$, and H=20.5. Due to potential source variability and different observing time, the U band data have significant uncertainty which has been taken into account in the subsequent analysis.

H-band observations were carried out in June 2000 using the Cambridge
InfraRed Survey Instrument (CIRSI) at the INT. The H filter is centred
at 1.65 $\mu$m with a width (FWHM) of 0.3 $\mu$m and a transmission peak
of 75\%. A total integration of 60 minutes was taken over the $30'
\times 30'$ used for the previous observations.

For several of our sources we also used fluxes in the Z-band taken with INT by Babbedge et al. (2004, in preparation, the Wide Field Survey).

All the images have been registered to a common World Coordinate System and resampled to a common pixel scale. This step allowed us to carry out the detection in one reference
image and the flux measurements in the others. This alleviates the
task of cross-correlating different catalogues and allows us to measure
magnitudes using the same aperture in all the images, as appropriate for SED analysis and required for the calculation of photometric redshifts. 
Source detection was performed using SExtractor (Bertin \& Arnouts
1996). The measured parameters include positions and positional errors, basic shape 
parameters as well as elongation and ellipticity, 
and different types of fluxes and magnitudes: isophotal, corrected
isophotal, fixed aperture and Kron-like automatic aperture as well as
errors. 
The automatic aperture magnitude is designed to give the most precise flux measure, and is the
one used as the magnitude in this work.  For the above mentioned reasons, we require the
magnitudes to be measured in the same aperture in each band.  The
object detection and aperture estimation is performed in the r band
images and then magnitudes are obtained on all the images using that
aperture. Since there are some X-ray objects which fall in the
gaps between r band chips, a second catalogue was created by taking as
reference the g band image.  
The absolute photometry was tied to the Wide Field Survey (optical) catalogue
already existent in that area and 2MASS (infrared) using
bright non saturated stars. 
85\% of the X-ray sources were found to have an optical counterpart, a fraction comparable to the detection rate achieved with Spitzer IRAC.

\subsection{Redshift Measurements}

Spectroscopic redshifts are available for only 13 of the 99 extragalactic sources in common between Chandra and Spitzer. These include 6 type-1 QSOs and two low-redshift galaxies observed by Gonzales-Solares et al. (2004, in preparation, including details on this optical spectroscopic run).
Additional optical spectroscopy has been obtained at Palomar with the COSMIC spectrograph on June 10 \& 11, 2004, providing the redshift for 5 further sources (2 type-1, 2 type-2 AGNs and a galaxy).

For the remaining objects, and thanks to the good coverage of the near- and mid-IR spectrum with the 4 IRAC channels, we estimated photometric redshifts with a tool
based on the Hyper-z code (Bolzonella et al. 2000), as discussed in detail by Polletta et al. (2004, in preparation). Hyper-z allowed a large database of galaxy spectra to be synthesized, including the effects of dust
extinction. The following parameter ranges have been explored with Hyper-z:
$A_V$ between 0 and 0.7, redshifts between 0 and 3.5, absolute magnitudes in the $g'$ band between -19 and -26.
As expected, for the typical SEDs of our sample objects, the redshift estimates turned out to be rather insensitive to the amount of extinction. 

Similar photometric redshift approaches have been successfully attempted for X-ray selected source populations (Franceschini et al. 2002; Zheng et al. 2004), in which cases a comparison with spectroscopic measurements has revealed fair accuracy in the procedure (errors $\delta z \sim 0.1$).
Independent photometric redshifts estimates for SWIRE galaxy samples are reported in Rowan-Robinson et al. (2004, AJ submitted; see also Rowan-Robinson 2003) and by Babbedge et al. (2004). Fair agreement between these codes is found by Babbedge et al., although in the current case the z-estimate is made more uncertain by the more complex nature of SEDs of AGN host galaxies.

Given the typical optical-IR SEDs of our sources, we expect that the photometric redshifts are more accurate for galaxy-dominated spectra and type-2 AGNs, which together make up the majority in our sample.
Obviously, for the typically flat and featureless SEDs of type-1 QSOs, they are expected to be much less reliable. This expectation is confirmed by a comparison of photometric and spectroscopic redshifts for our 13 spectroscopically identified sources.

\section{RESULTS }\label{sec:results}

\subsection{Cross-Correlation Analysis and Source Identification}\label{3.1}

\begin{figure}
 \includegraphics[width=0.5\textwidth]{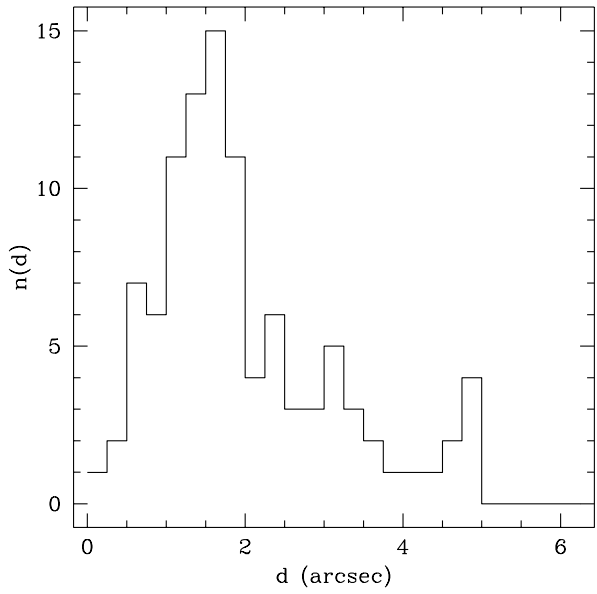}
\caption{The distribution of positional distances between the centroids of the 3.6 $\mu$m sources and the Chandra sources. The distribution includes a slight systematic offset in declination, also apparent in Fig. \ref{fig2}, which however does not influence the source identification.
}
\label{fig3}
\end{figure}

A simple near-neighbour search has been performed to cross-correlate the Spitzer and Chandra source catalogues within the Chandra ACIS-I chip image, using a $d=5^{\prime\prime}$ search radius (roughly the quadratic sum of the astrometric errors).
The distribution of separations between the Chandra and Spitzer centroids is reported in Fig. \ref{fig3}.

The Spitzer source catalogue used for the cross-correlation was a subset of the full band-merged catalogue for the IRAC and MIPS 24 $\mu$m bands. It contained only those sources with a signal-to-noise greater than 4 in at least 2 of the Spitzer bands. This catalogue contained 3420 objects within the region of the Chandra ACIS-I image, compared with the 122 Chandra full-band detections.

Altogether, we find reliably associated counterparts for 102 of the 122 Chandra sources (84\% in total). The vast majority of these are detected with the IRAC channels 1 and 2 (3.6 and 4.5 $\mu$m): 100 of the 122 Chandra sources in each case. Such an identification fraction decreases when considering the longer wavelength IRAC channels or MIPS. 59 Chandra objects are reliably associated with MIPS 24 $\mu$m sources (all of them having IRAC counterparts) and just 1 had a MIPS 70 $\mu$m counterpart. More details on the source identification statistics are reported in Table 1, where the first column indicates the Spitzer band, the second the number of Spitzer sources detected above the 4$-\sigma$ limiting flux $S_{lim}$ (reported in the third column), the number of Chandra sources identified in the various channels, and their percentage over the total number of Spitzer sources.

To estimate the reality of the associations we have calculated the probability of random matches between the X-ray sources and the possible Spitzer counterparts. As discussed in Fadda et al. (2002), we assume the IR population to follow a Poisson spatial distribution, such as the probability P of a random association is
\begin{equation}
P = 1 - e^{-N \pi d^2}
\end{equation}
where $d$ is the offset distance between the X-ray and Spitzer source and $N$ is the areal number density of possible Spitzer counterpart derived from Fig. \ref{fig1}. 
The probability $P$ turned out to be sufficiently small to guarantee that almost all our associations are real. Approximately 90\% have $P<0.03$. Summing the probabilities we expect 2 false associations in the 102 matched sources.

Only 3 Chandra sources have multiple associations within $5^{\prime\prime}$ in the IRAC images. Of these, two (N1\_8 \& N1\_81) have been unambiguously identified as they lie near to the centre of the Chandra image where the Chandra PSF is small. Since the positional accuracy is dominated by the size of the Chandra PSF, we can reliably reject any associations with offsets greater than the $3\sigma$ size of the Chandra PSF for these sources, leaving only one association in each case. The third source (N1\_104) is near to the edge of the Chandra image and therefore has a large Chandra PSF which contains 2 Spitzer sources (as well as 2 optical sources) within the $3\sigma$ PSF extent. We assume the association to be the Spitzer source closest to the centre of the Chandra PSF and closest to the central optical source. This appears justified by the observed Seyfert-type spectrum found for this object (see Fig. \ref{figSED8}).

Of the 102 Spitzer-identified Chandra sources, 3 turned out to correspond to galactic stars based on position on colour-magnitude plots and on optical morphology, and are excluded from our subsequent analysis.

\begin{figure}
 \includegraphics[width=0.8\textwidth]{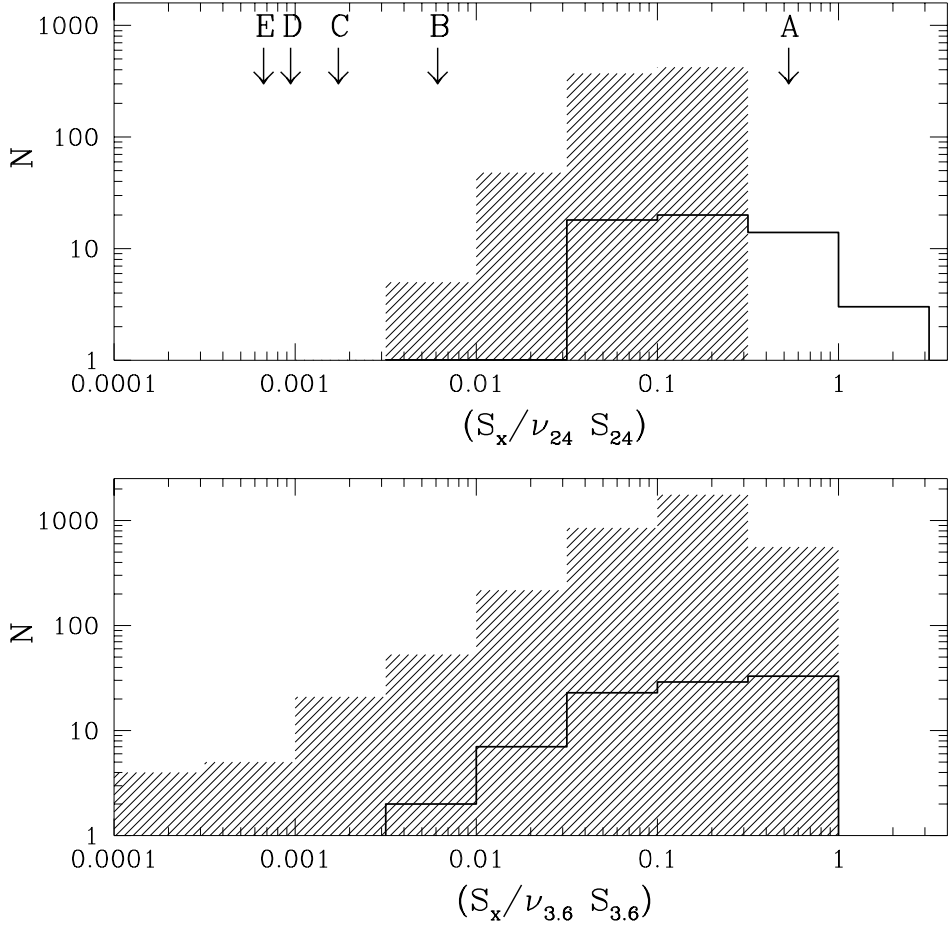}
\caption{Top: histograms of the ratio of the 0.5-8 keV band X-ray flux (S$_X$) to the 24 $\mu$m flux for sources detected by SWIRE in the area covered by Chandra. The continuous-line histogram refers to sources with Chandra counterparts, while the shaded histogram is the distribution of upper limits for those without Chandra associations. 
Bottom: same as top panel for the IRAC 3.6 $\mu$m sources.
Our adopted flux limits for the SWIRE samples are 5 $\mu Jy$ and 100 $\mu Jy$ for 3.6 and 24 $\mu$m respectively. 
In the top panel the arrows indicate the values of representative active galaxies: A: NGC 4151, B: NGC 6240, C: IRAS19254s, D: M82, and E: Mkn 231.
}
\label{histo}
\end{figure}

We report in Fig. \ref{histo} histograms of the ratio of the X-ray flux (in the total 0.5-8 keV band) to the 3.6 and 24 $\mu$m fluxes for sources detected by SWIRE in the Chandra area brighter than 5 $\mu Jy$ and 100 $\mu Jy$, respectively. 
The figures compare the flux ratio distributions for sources with Chandra counterparts (continuous-line histograms) and those without Chandra associations (shaded histograms). The latter are the distribution of the upper limits on the flux ratios computed using the Chandra 3$\sigma$ X-ray sensitivity limits as in Sect. \ref{X}.                                               

The arrows in the top panel of Fig. \ref{histo} indicate the observed values of the flux ratios for representative active galaxies: a type-1 (A: NGC 4151) and type-2 AGNs (B: NGC 6240, C: IRAS19254s), an ultraluminous IR quasar with high X-ray absorption (E: Mkn 231), and a low-luminosity starburst (D: M82).  For these sources, the 24 $\mu$m flux ($\nu_{24}S_{24}$) is estimated from the 25 $\mu$m IRAS data, while the X-ray data are taken from George et al. (1998), Cappi et al. (1999), Franceschini et al. (2003) and Braito et al. (2004).  X-ray to IR flux ratios in our Chandra sample are then typical of X-ray luminous AGNs, and definitely larger than those of starburst galaxies or even heavily absorbed type-2 QSOs.

\subsection{Analysis of the Spectral Energy Distributions }\label{3.2}

The positions, Spitzer IRAC and MIPS data, X-ray fluxes, spectral classification, the X-ray hardness ratio and luminosities for our SWIRE-Chandra sample are reported in Table \ref{seds}.
The excellent UV-optical-IR spectral coverage available for our sources has prompted us to perform a very extensive analysis based on the Hyper-z code (Bolzonella et al. 2000) and a wide variety of additional spectral templates, aimed at a physical characterization of the sources. 

Basically, starburst-dominated and quasar-dominated sources can be discriminated on the basis of the presence or absence of the classical UV-excess and flat optical SED in type-1 QSOs, or from the mid-IR excess characterizing the spectra of both type-1 and type-2 AGNs. The latter is due to emission of very hot dust with temperatures close to the grain sublimation, and essentially missing in purely starburst regions. We defer to Berta et al. (2003) and Prouton et al. (2004) for further discussion about this point.
The diagnostic power offered by full UV-optical-IR spectral coverage for disentangling AGNs and starbursts may be appreciated by considering that the typical peak around 1 $\mu$m of the integrated emission of stellar populations in normal galaxies roughly (and coincidentally) corresponds to a minimum of the spectrum of quasar nuclei between the UV "big bump" and the near-IR hot dust component.   We will further discuss this point in subsequent Sections.

While an exhaustive description of the spectral database used for our analysis can be found in Polletta et al. (2004 in preparation), we summarize here its main features. It consists of 22 spectral templates which are derived from the observed SEDs of various galaxy classes, while in the optical/near-IR the observations are complemented with a spectrophotometric synthesis code presented in Berta et al. (2004). The library contains spectral templates of all kinds of AGNs, from optical quasars (Brotherton et al. 2001), to red quasars (Gregg et al. 2002), Seyfert 1 to 2 galaxies, combined quasar/ULIRG sources like the type-1 QSO Mkn 231 and the type-2 quasar/ULIRG Superantennae South (IRAS19254s, Berta et al. 2003). 

\begin{figure}
 \includegraphics[width=0.95\textwidth]{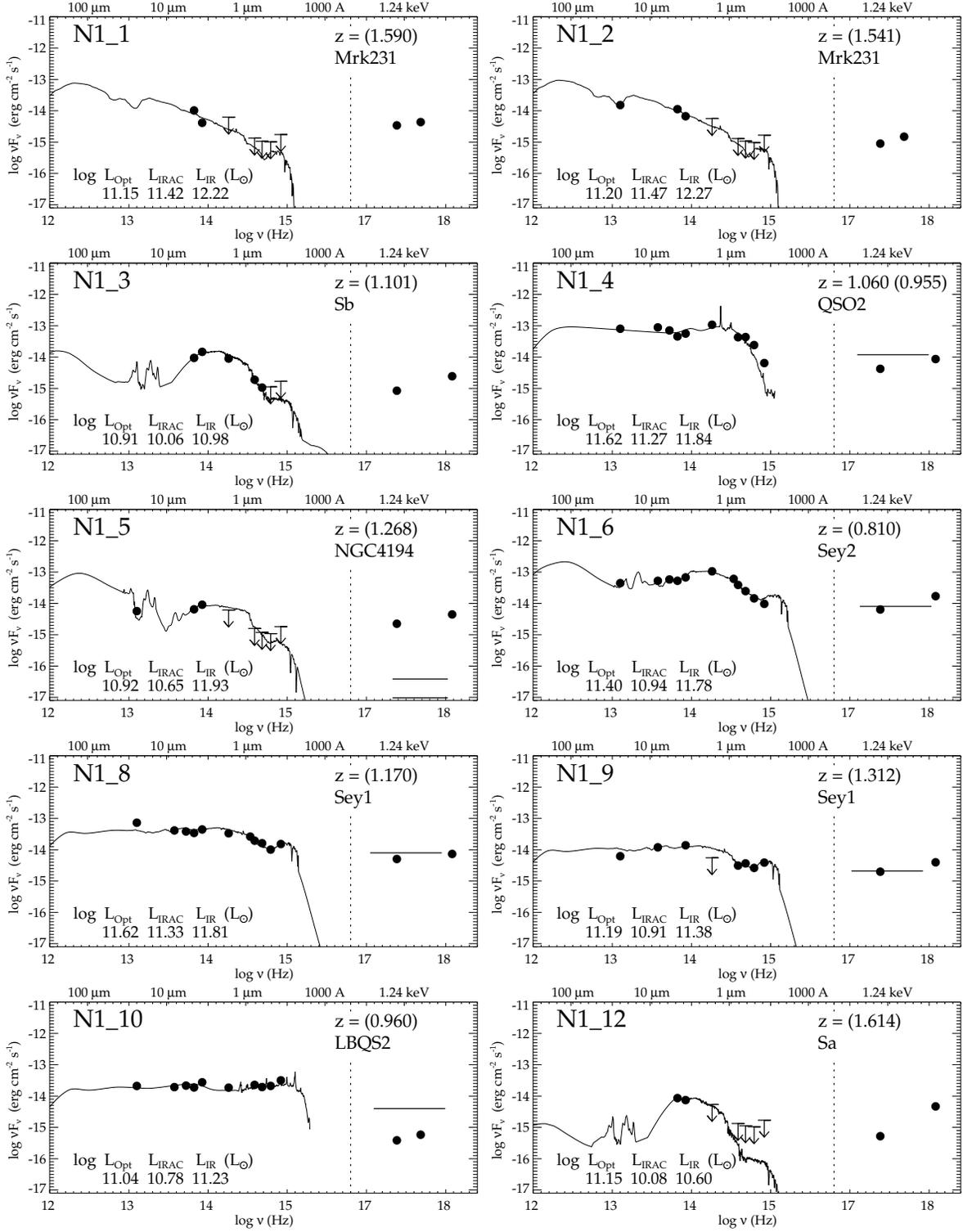}
\caption{Broad band Spectral Energy Distributions and spectral best-fits for all 99 Chandra/SWIRE sources. The scale on the X-axis has been artificially broken between the UV and X-ray (dotted line) for plot optimization. See text for more detailed explanation.
}
\label{figSED1}
\end{figure}

\begin{figure}
 \includegraphics[width=0.95\textwidth]{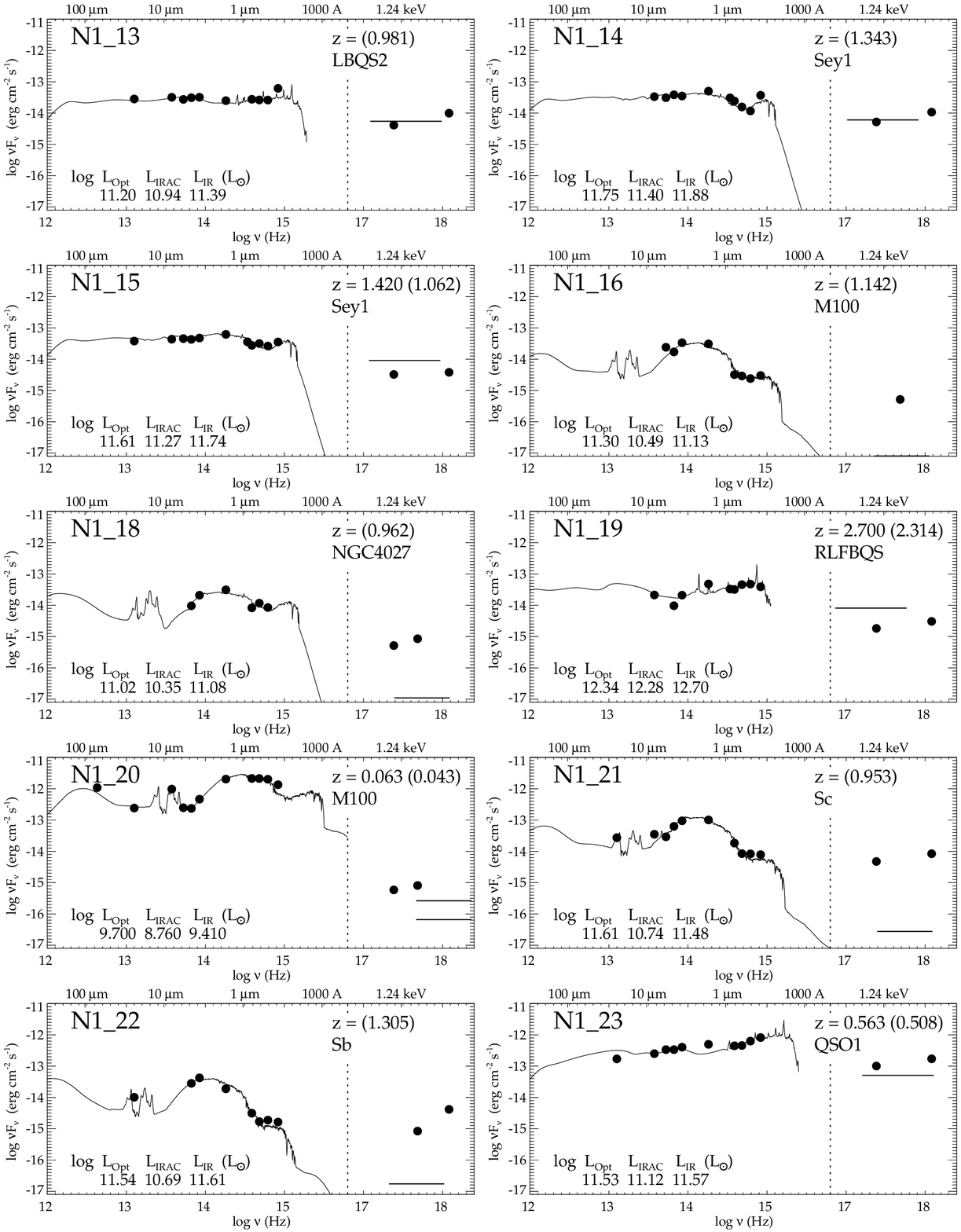}
\caption{Broad band Spectral Energy Distributions and spectral best-fits (continued).
}
\label{figSED2}
\end{figure}

\begin{figure}
 \includegraphics[width=0.95\textwidth]{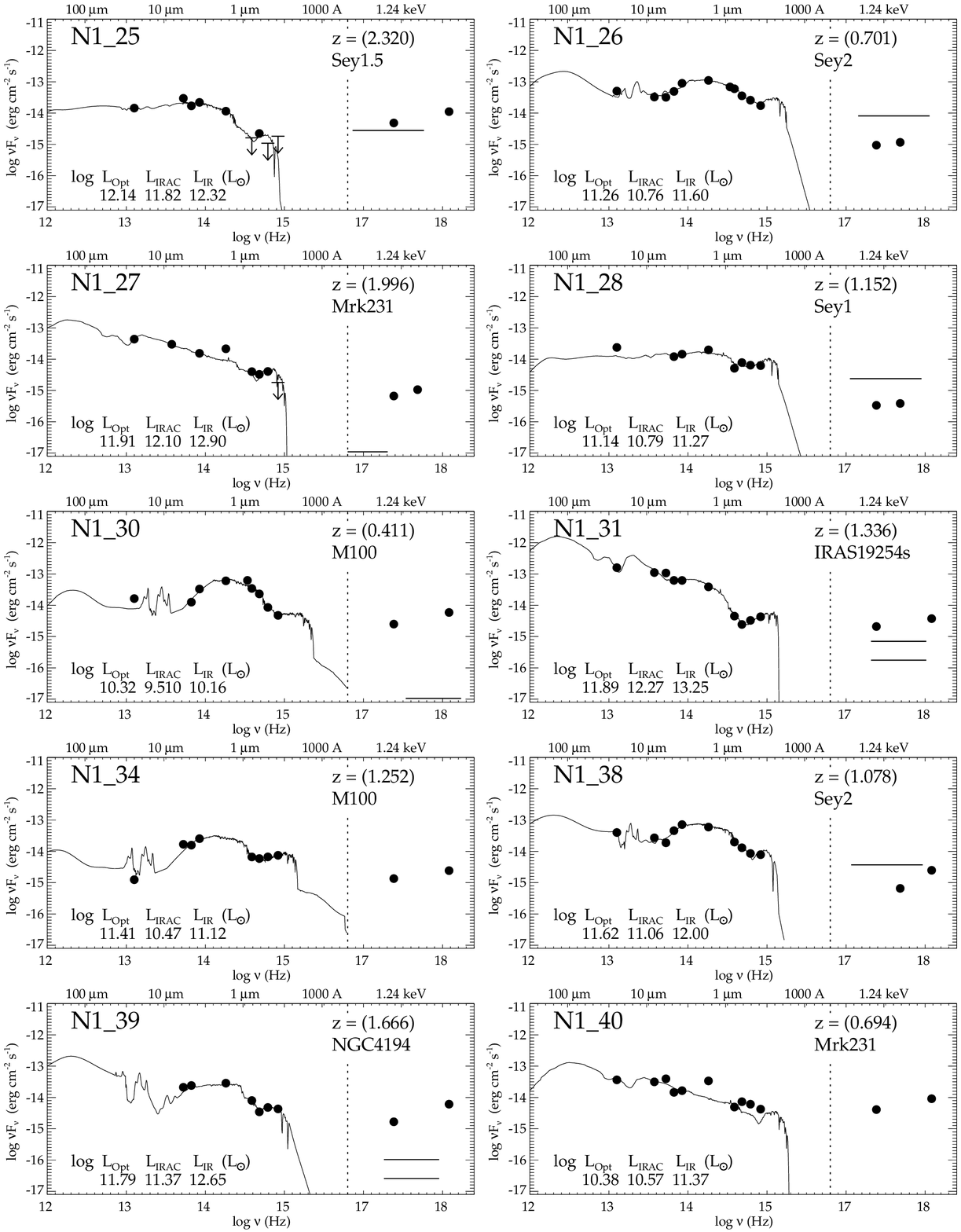}
\caption{Broad band Spectral Energy Distributions and spectral best-fits (continued).
}
\label{figSED3}
\end{figure}

\begin{figure}
 \includegraphics[width=0.95\textwidth]{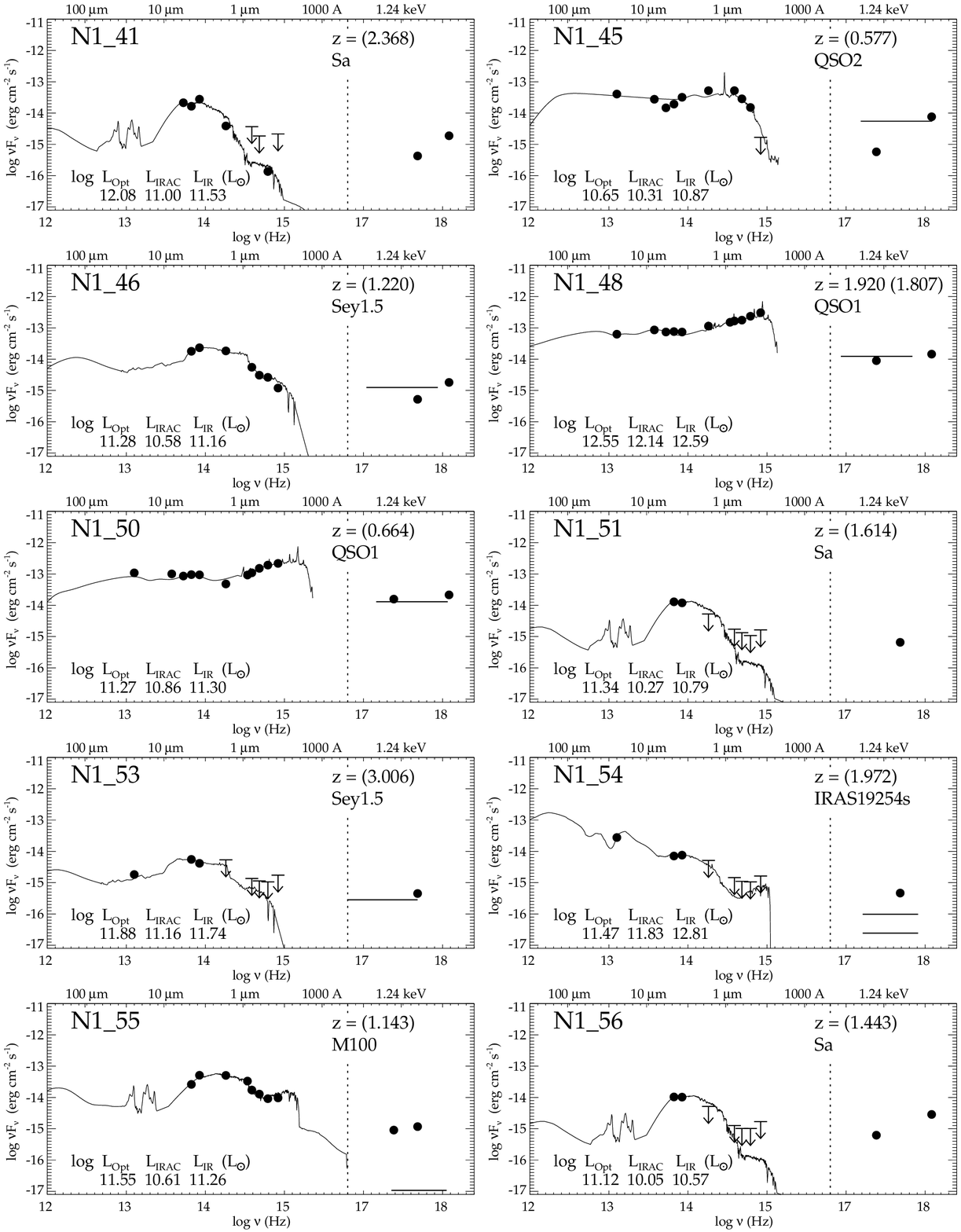}
\caption{Broad band Spectral Energy Distributions and spectral best-fits (continued).
}
\label{figSED4}
\end{figure}

\begin{figure}
 \includegraphics[width=0.95\textwidth]{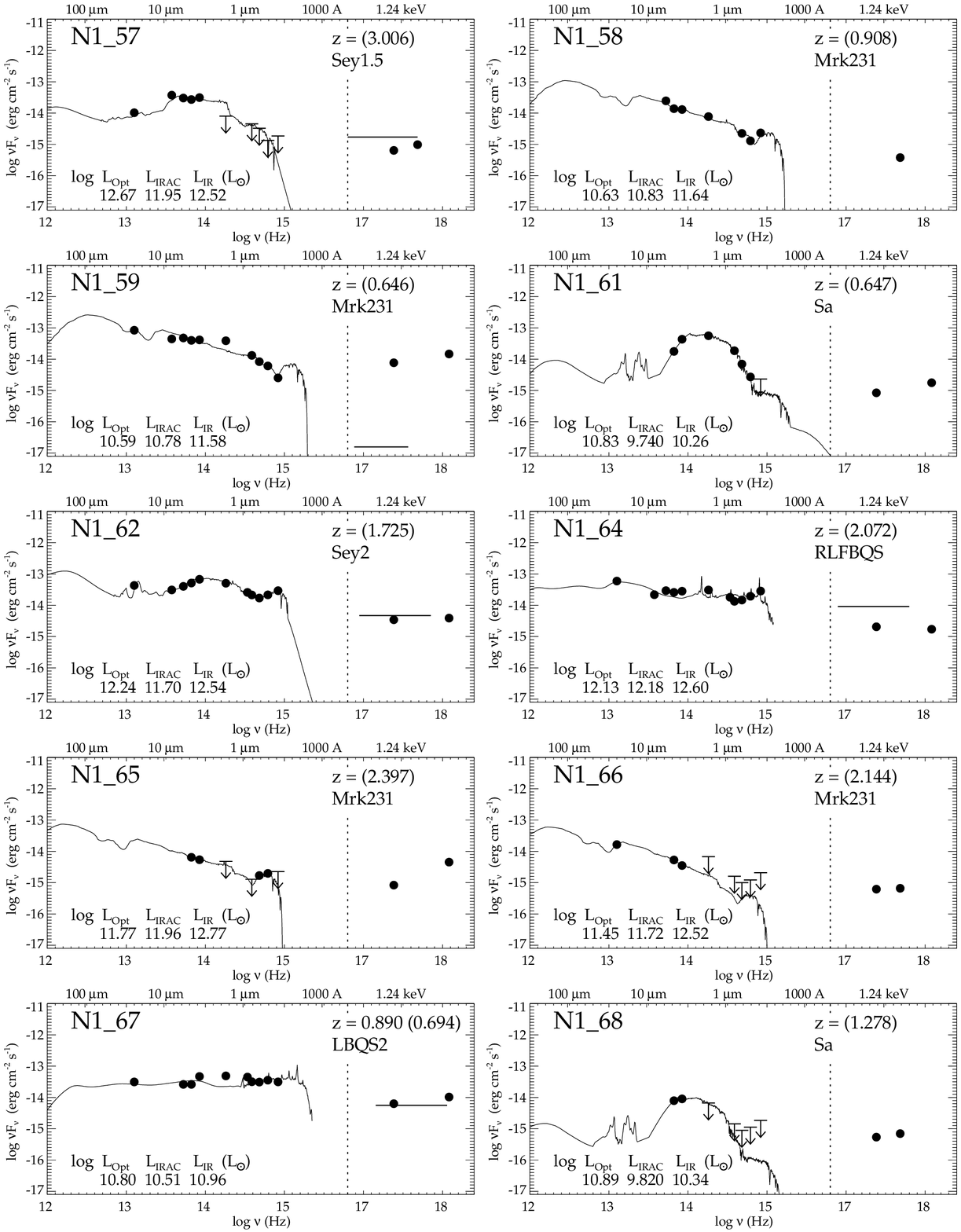}
\caption{Broad band Spectral Energy Distributions and spectral best-fits (continued).
}
\label{figSED5}
\end{figure}

\begin{figure}
 \includegraphics[width=0.95\textwidth]{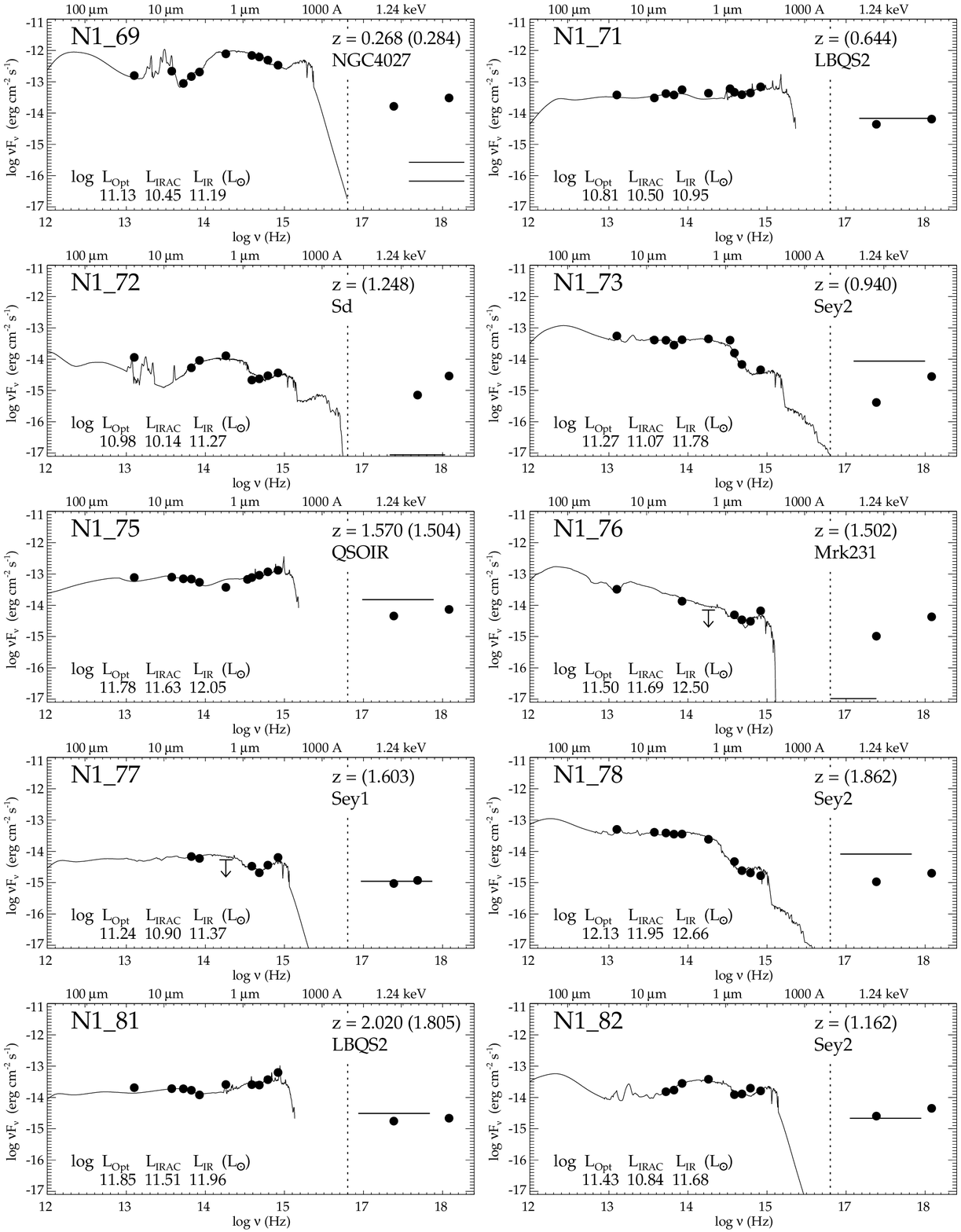}
\caption{Broad band Spectral Energy Distributions and spectral best-fits (continued).
}
\label{figSED6}
\end{figure}

\begin{figure}
 \includegraphics[width=0.95\textwidth]{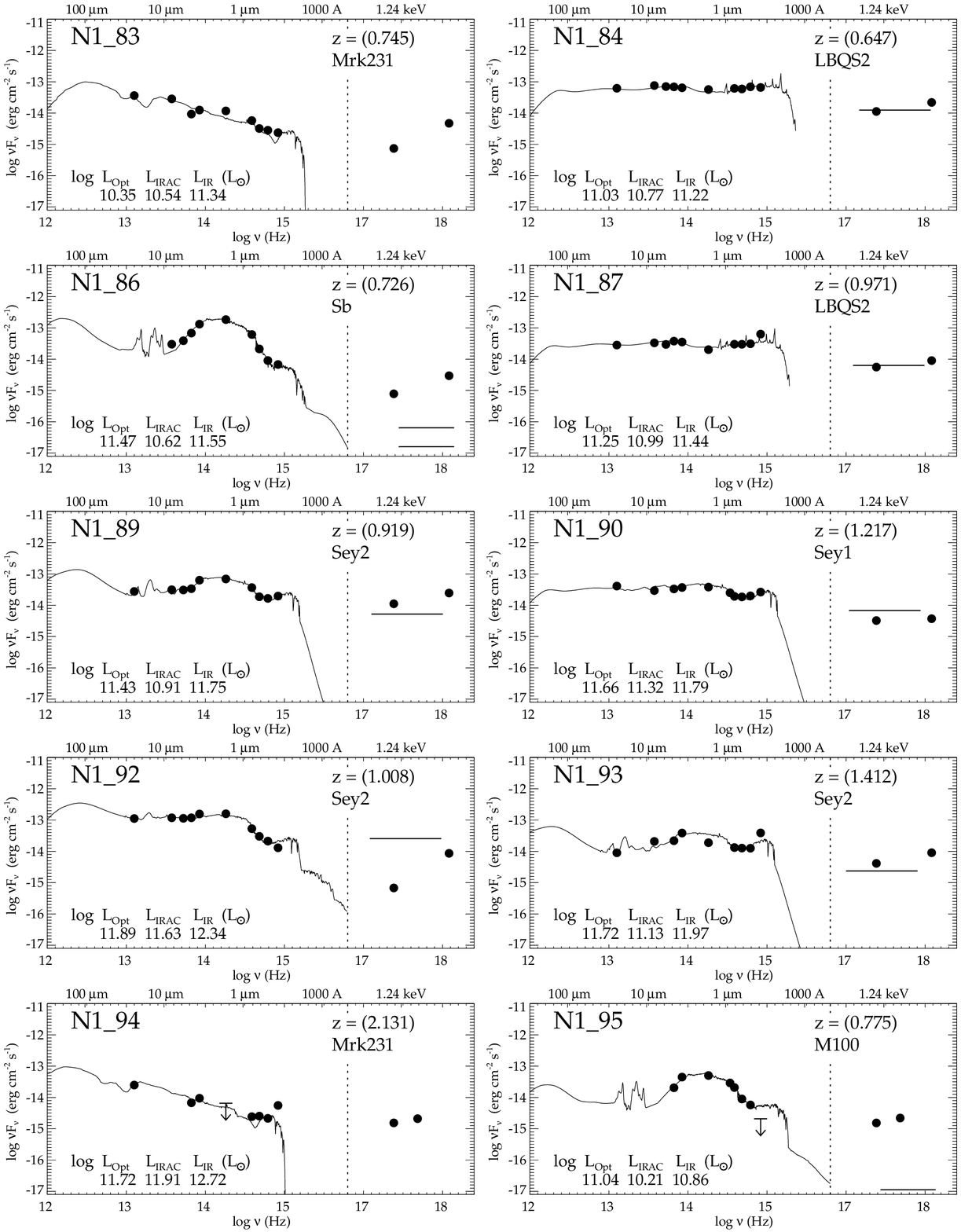}
\caption{Broad band Spectral Energy Distributions and spectral best-fits (continued).
}
\label{figSED7}
\end{figure}

\begin{figure}
 \includegraphics[width=0.95\textwidth]{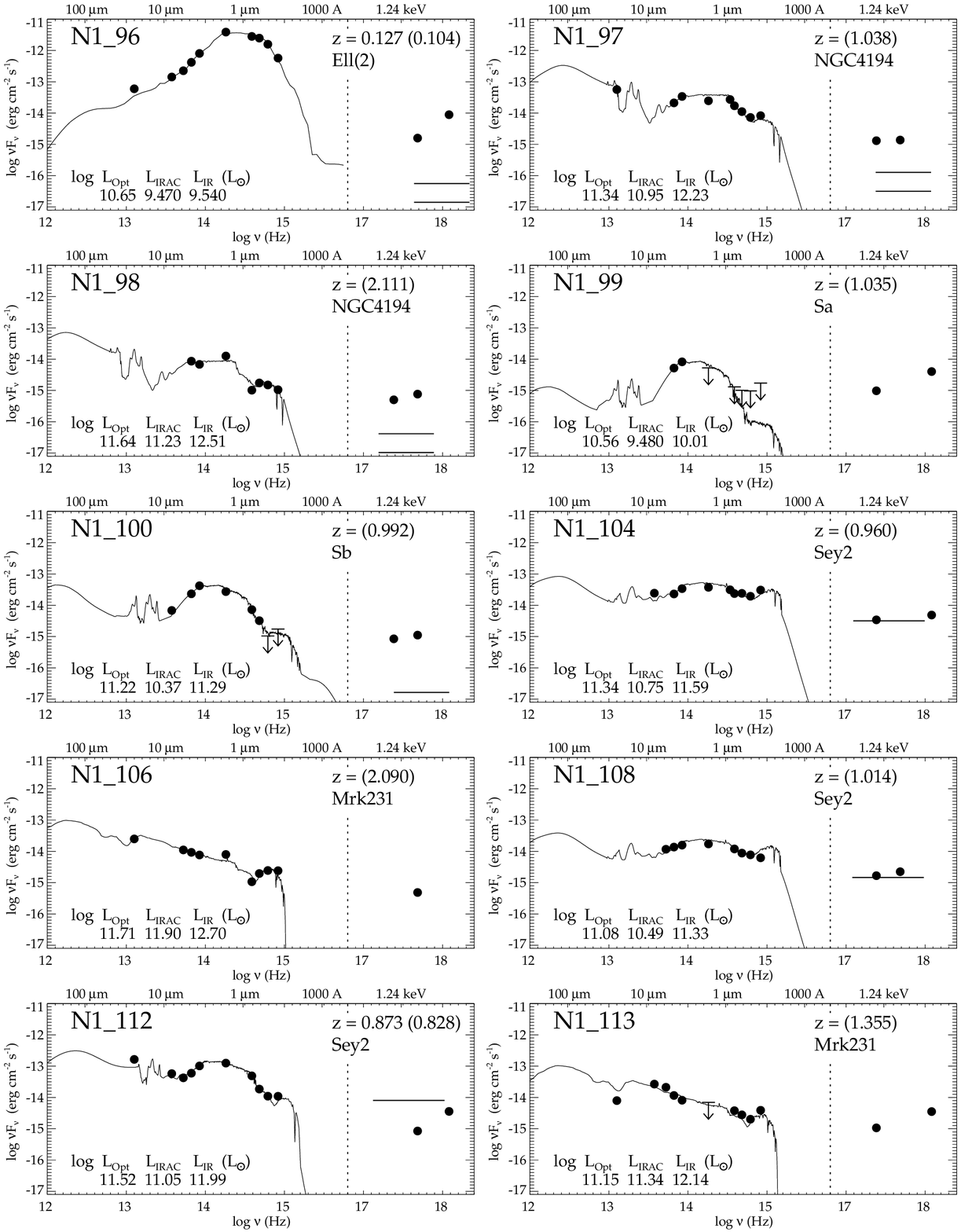}
\caption{Broad band Spectral Energy Distributions and spectral best-fits (continued).
}
\label{figSED8}
\end{figure}

\begin{figure}
 \includegraphics[width=0.95\textwidth]{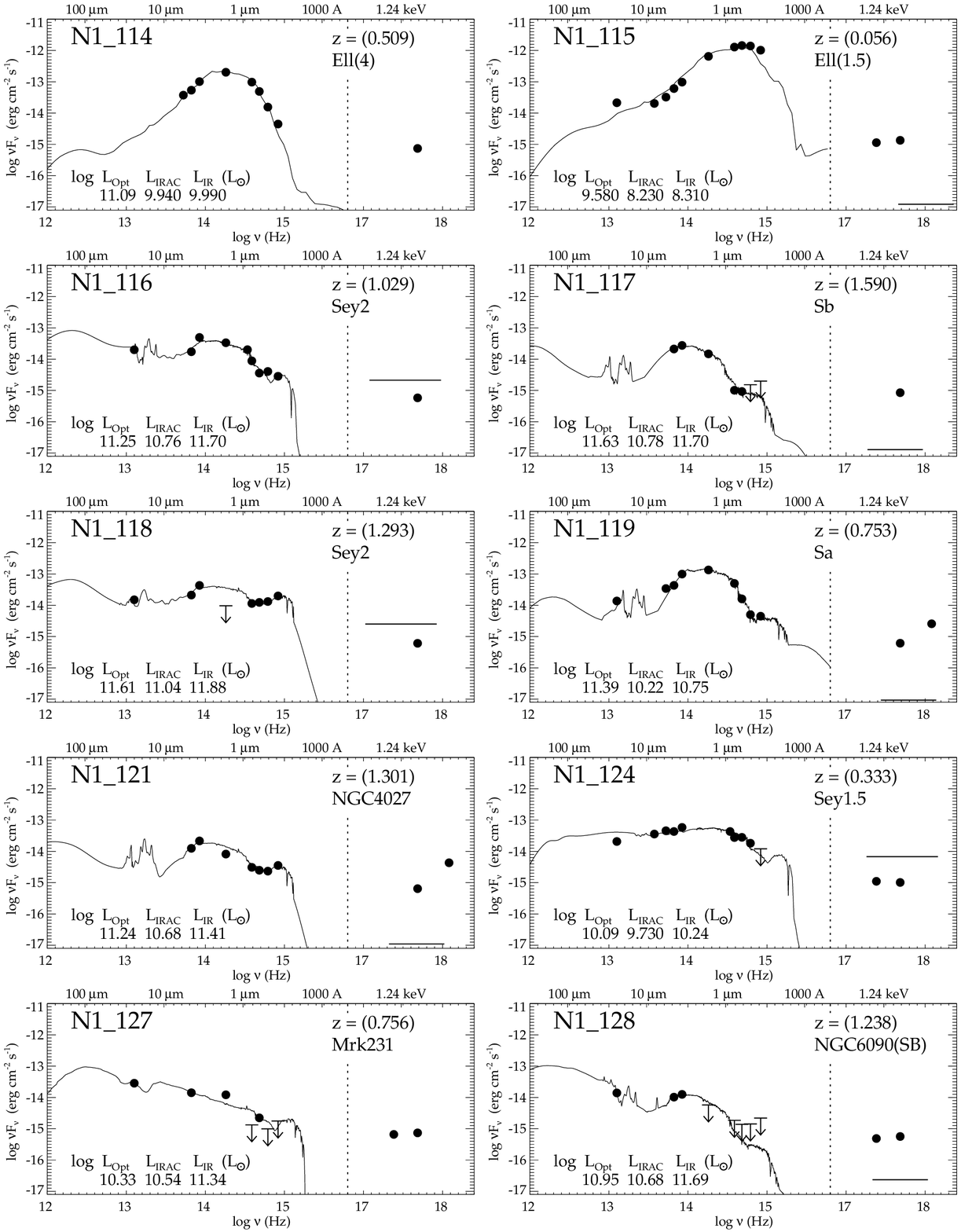}
\caption{Broad band Spectral Energy Distributions and spectral best-fits (continued).
}
\label{figSED9}
\end{figure}

\begin{figure}
 \includegraphics[width=0.95\textwidth]{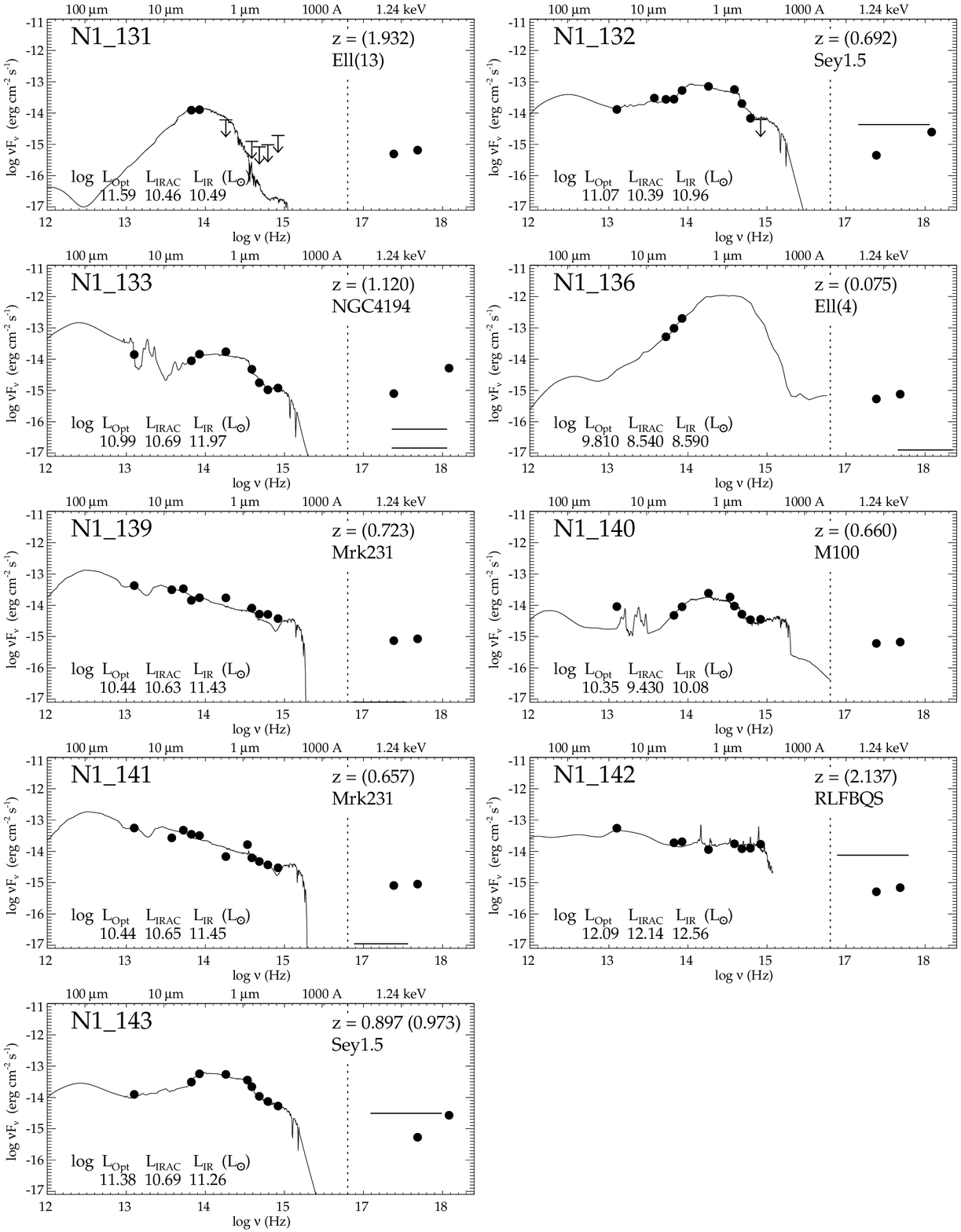}
\caption{Broad band Spectral Energy Distributions and spectral best-fits (continued).
}
\label{figSED10}
\end{figure}

Figures \ref{figSED1} to \ref{figSED10} display the Spectral Energy Distributions and spectral best-fits for the 99 extragalactic Chandra/SWIRE sources in our sample (3 stars are excluded). Data points are plotted for the 4 IRAC bands (3.6$\mu$m, 4.5$\mu$m, 5.8$\mu$m \& 8$\mu$m), MIPS 24$\mu$m band,
U, g', r', i', Z, H, and Chandra X-ray bands. Where the sources are detected in the soft (0.5 - 2 keV) or hard (2 - 8 keV) Chandra images, their fluxes are converted to flux densities at 1 and 5 keV respectively. Where the Chandra source is not detected in both bands, the full band (0.5 - 8 keV) flux is plotted, corrected to a flux density at 2 keV.

Spectral fits are also shown in the figures, where the best-fitting spectral type is given in the upper-right corner of each plot, together with the photometric redshift (value given in brackets).
The spectroscopic redshifts available for 13 sources are reported outside brackets. Also given in each panels are values for the bolometric luminosities within the observed optical (0.3 - 3 $\mu$m), IR (3 - 1000 $\mu$m), and IRAC (3 - 10 $\mu$m) bands for the best-fitting spectral type.

The spectral fitting was performed only on the optical and infrared data points. We have therefore used the best-fitting spectral types to make a prediction for the corresponding X-ray emission in each cases. For those spectral types where the infrared emission is dominated by star-formation processes, we have calculated the corresponding X-ray emission based on our estimated star-formation rate (SFR). Following Franceschini {\it et al.} (2003), the SFR is calculated from the bolometric infrared luminosity. Given the SFR, the number of High-Mass X-ray Binaries (HMXBs) may be inferred along with their corresponding 2 - 10 keV emission. For the case of ULIRGs not dominated by an AGN, it is assumed that all of the X-ray emission comes from HMXBs. In the case of lower-luminosity starbursts, a quarter of the 2 - 10 keV luminosity may be due to HMXBs (the rest being attributable to LMXBs and the quiescent disc). Therefore our expected 2 - 10 keV luminosity due to star-formation is given by:
\begin{equation}
L_{2-10 keV} = \left(\frac{1}{f}\right) \frac{L_{IR}}{2.2\times10^4}
\end{equation}
where $f$ ranges from 0.25 to 1 (Franceschini {\it et al.} 2003). These expected values of X-ray flux are plotted for each of the starburst-like and ``normal galaxy'' spectral types. In all cases they are found to under-predict the observed X-ray flux providing clear evidence for the existence of a hidden AGN in these sources.

For those spectral types where the infrared emission is dominated by the AGN (Seyfert types 1 \& 2 and QSOs), we have predicted the associated X-ray emission based on the mean radio-quiet quasar spectrum of Elvis et al. (1994). This is achieved based on the ratio of the flux density at 4.5 $\mu$m to that at 2 keV: in roughly 2/3 of cases this provides a good prediction of the observed X-ray flux. In the remaining cases the X-ray emission is over-predicted, indicating a component of the infrared emission not originating from the AGN.

For the spectral types best-fitted by the Mkn 231 template (15 sources in total), the actual X-ray flux of Mkn 231 is plotted as observed by Braito et al. (2004). This strongly under-predicts the observed X-ray fluxes in all cases, consistent with the very low ratio of $S_x/S_{3.6}$ of Mkn 231 indicated in Fig. \ref{histo} ("D" arrow). Our SWIRE-Chandra objects display typical X-ray energy fluxes that are within $\sim$ a decade of those measured in the infrared, which demonstrates an AGN dominance in these sources.

\subsection{Source Classification}\label{3.2.2}

Our analysis has found good spectral solutions for a large majority of the sources. 
Figures \ref{figSED1} to \ref{figSED10} illustrate the quality of the observational spectral coverage and that of our spectral best fits. For a few sources in the figure the photometric redshift is compared with the available spectroscopic measures, with a fairly good agreement between the two.

For 7 sources the observational SED coverage is not enough for reliable source characterization. In these cases our adopted classification exploits additional constraints from the X-ray spectrum and the IR to X-ray flux ratios.

We have then grouped sources according to the following four categories: 1) type-1 quasars and AGNs, including 39 objects out of 99 (39\%); 2) type-2 quasars and AGNs (23 of 99, 23\%); normal and starburst galaxies (33 objects, 33\%); old passive galaxies with purely stellar mid-IR spectra (5 galaxies, 5\%).
We will examine in the next Section properties of these various classes.

\subsection{A New Diagnostic Diagram based on Optical-IR SEDs}

We have first attempted to identify a diagnostic diagram able to disentangle various categories of high-redshift active galaxies based on optical-IR colors, which are easily achievable for huge numbers of sources from combined deep surveys with Spitzer and optical ground-based telescopes. 

\begin{figure}
 \epsfig{file=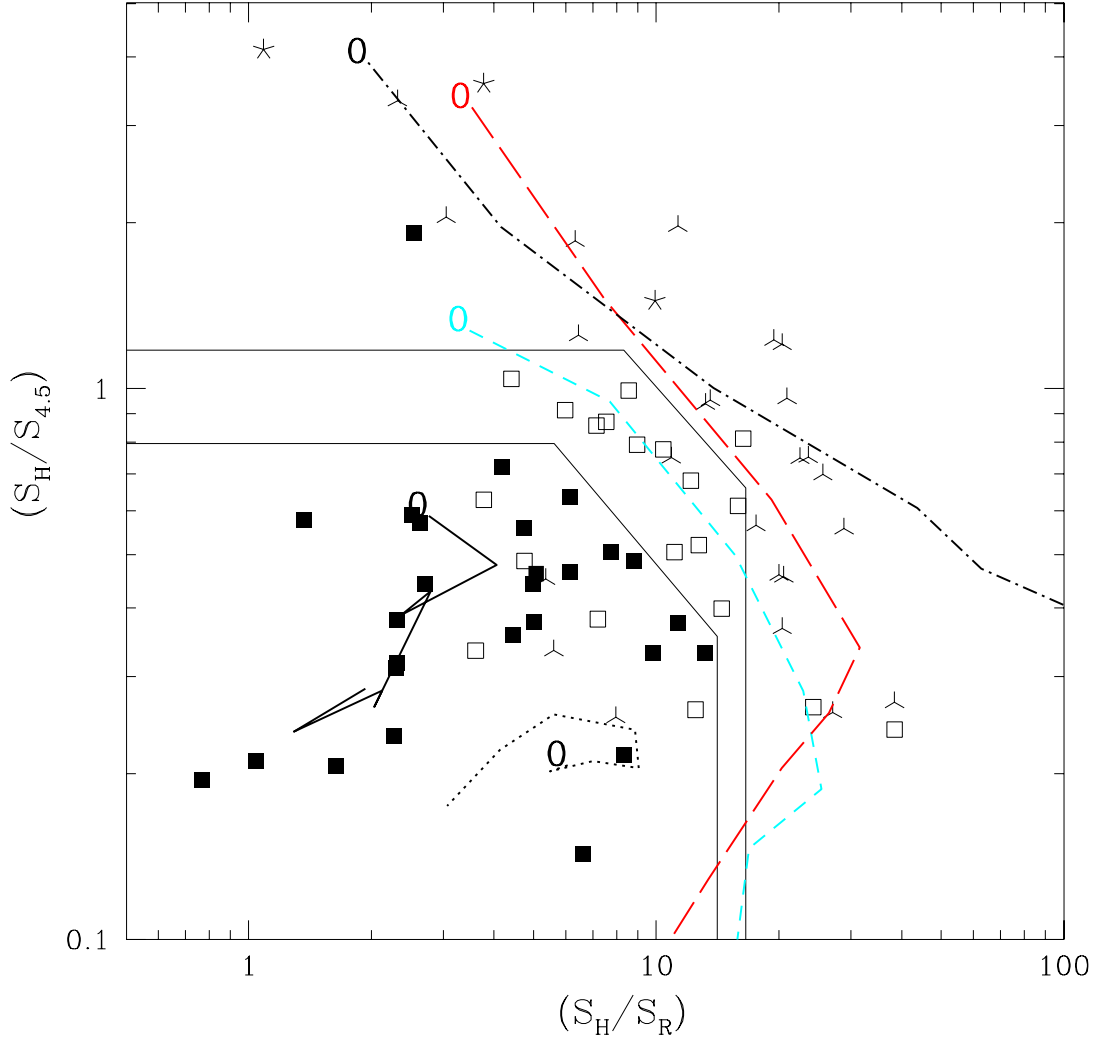,width=0.9\textwidth,clip=} 
\caption{Color-color plot with $S_H/S_{4.5}$  versus $S_H/S_R$. $S_H$ and $S_R$ are the flux densities in the near-IR H-band and the optical R-band. Four source categories are here reported: type-1 quasars and AGNs (filled squares); type-2 AGNs (open squares); normal spirals and starbursts (3-leg stars), early-type galaxies (5-leg stars). 
These four categories preferentially populate different regions of the diagram, which are indicated by the two thin broken lines. The lower line is defined by the two points on the graph at ($10^{0.75}$; $10^{-0.1}$) and  ($10^{1.15}$; $10^{-0.45}$). The upper line is defined by the two points at ($10^{0.92}$; $10^{0.07}$) and  ($10^{1.22}$; $10^{-0.18}$). 
The other thicker lines show the expected evolution of colors as a function of redshift for various 
spectral templates: (black) dot-dashed line: elliptical galaxy; (red) long-dash: Sb spiral/starburst; (cyan) short-dash: Seyfert-2; (black) continuous line: Seyfert-1 or quasar; (blue) dotted line: Mkn 231. The color for z=0 is the point marked with "0"; then lines are drawn from here to the points corresponding to z=0.5, 1, 1.5, 2, 2.5, 3 and 3.5.
}
\label{diagno}
\end{figure}

We have found that a maximum segregation of our 4 categories is obtained by combining the H-band to 4.5 $\mu$m color against the H-band ($\lambda_{eff}=1.7\ \mu$m) to R-band magnitudes. As illustrated in Fig. \ref{diagno}, our three classes tend to populate different regions of the color-color diagram which are delimited by the two continuous think broken lines. In this plot, type-1 AGNs occupy the lower left corner, corresponding to sources with excess 4.5 $\mu$m emission (of quasar origin) but blue or very blue optical to H-band spectra. With respect to these, type-2 AGNs appear to be shifted towards an excess in the 4.5 $\mu$m and H-band fluxes. Finally, galaxies classified as normal (i.e. starbursts or quiescent disks) tend to occupy a quite complementary domain of the graph with respect to the previous two: if such galaxies are relatively blue in the H to optical color, they show no mid-IR excess (upper left corner); if they are red in the optical, they are also red in the near- to mid-IR.

This behaviour is easily understandable based on what we expect the typical source SEDs to be and anticipated in Sect. \ref{3.2.2}: type-1 quasars have "concave" SEDs with a minimum in the rest-frame around 1 $\mu$m. Normal galaxies have a maximum around 1 $\mu$m being redshifted towards longer wavelengths, which is a quite complementary spectral behaviour compared to quasars. Type-2 AGNs have spectra similar to those of type-1, but with an excess in the H-band corresponding to the emergent contribution of the host galaxy. Another feature of type-2 AGNs is that they lack the UV-excess of type-1 objects and avoid the region of very blue H-R colors (low values of the $S_H/S_R$ ratio).

The fluxes in the H-band and IRAC 4.5 $\mu$m are particularly critical to separate the contributions of the host galaxy from that of the nuclear AGNs. However our result here indicates that only a two-color diagnostic may work, whereas a single-color criterion (e.g. that based on the $S_H/S_{4.5}$ color) is subject to high degeneracy.

Note that the diagram of Fig. \ref{diagno} seems to work for sources within our properly sampled redshift interval, i.e. up to $z\simeq 2.5$, but with poor statistics above $z=1.5$. A more extensive dataset will be needed to prove if a similar diagnostics might work on a wider redshift interval.

We have attempted to verify the effect of the source redshift in this plot in a modellistic way, by overplotting lines corresponding to the expected evolution of colors as a function of redshift, for various different spectral templates. For each of the overplotted lines in Fig. \ref{diagno}, the color for redshift z=0 is in correspondence of the start point marked with "0". Then lines are drawn from here to the points corresponding to z=0.5, 1, 1.5, 2, 2.5, 3 and finally 3.5. The (black) dot-dashed line is for an elliptical galaxy template, the (red) long-dash for a Sb spiral/starburst, the (cyan) short-dash for a Seyfert-2, the (black) continuous line for a type-1 Seyfert or quasar, and finally the (blue) dotted for the Mkn 231 spectral template.
It is evident that, while for typical type-1 AGNs (the latter two) the evolution of colors with redshift is very modest and they keep well within the lower left quadrant for any z (due to the featureless broad-band spectra), strong color evolution is expected for all other categories. In particular, the H-R-4.5$\mu$m colors migrate from the upper left to lower right at increasing redshifts.

In any case the expected colors remain within the boundaries for the various categories up to z=3.5, except for type-2 AGNs and starbursts for which they tend to confuse at $z>1.5-2$.

\subsection{Flux Correlations and Color-Color Plots}

The relation between X-ray and mid-IR fluxes is an important one to constrain the distribution of the circumnuclear absorbing medium and eventually to test the unification scheme. Various correlation analyses have been published (Maiolino et al. 1995; Krabbe et al. 2001; Lutz et al. 2004), not all of them finding unequivocal results, however.

\begin{figure}
 \begin{tabular}{cc}
 \psfig{file=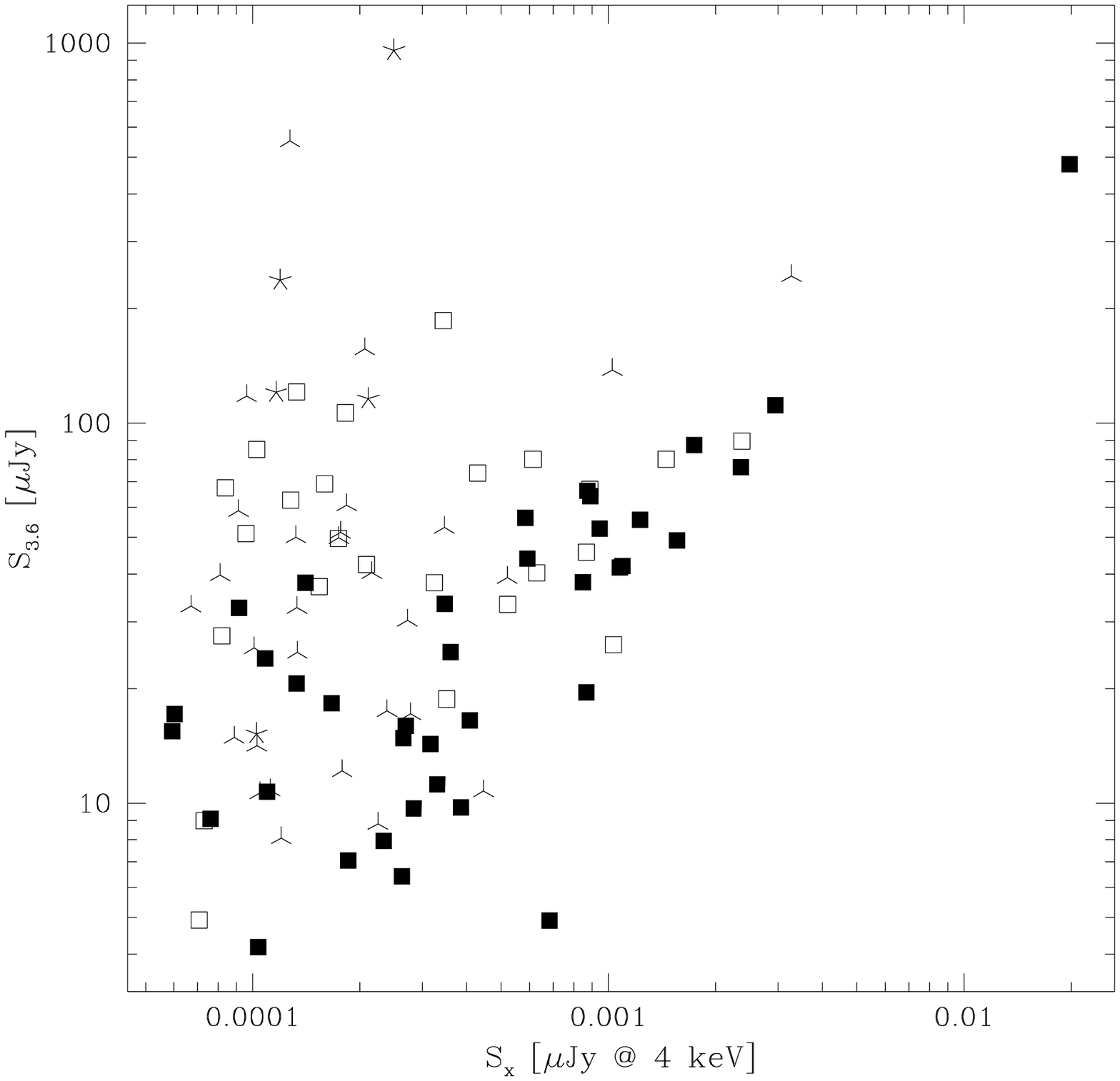,width=0.5\textwidth,clip=} &  \epsfig{file=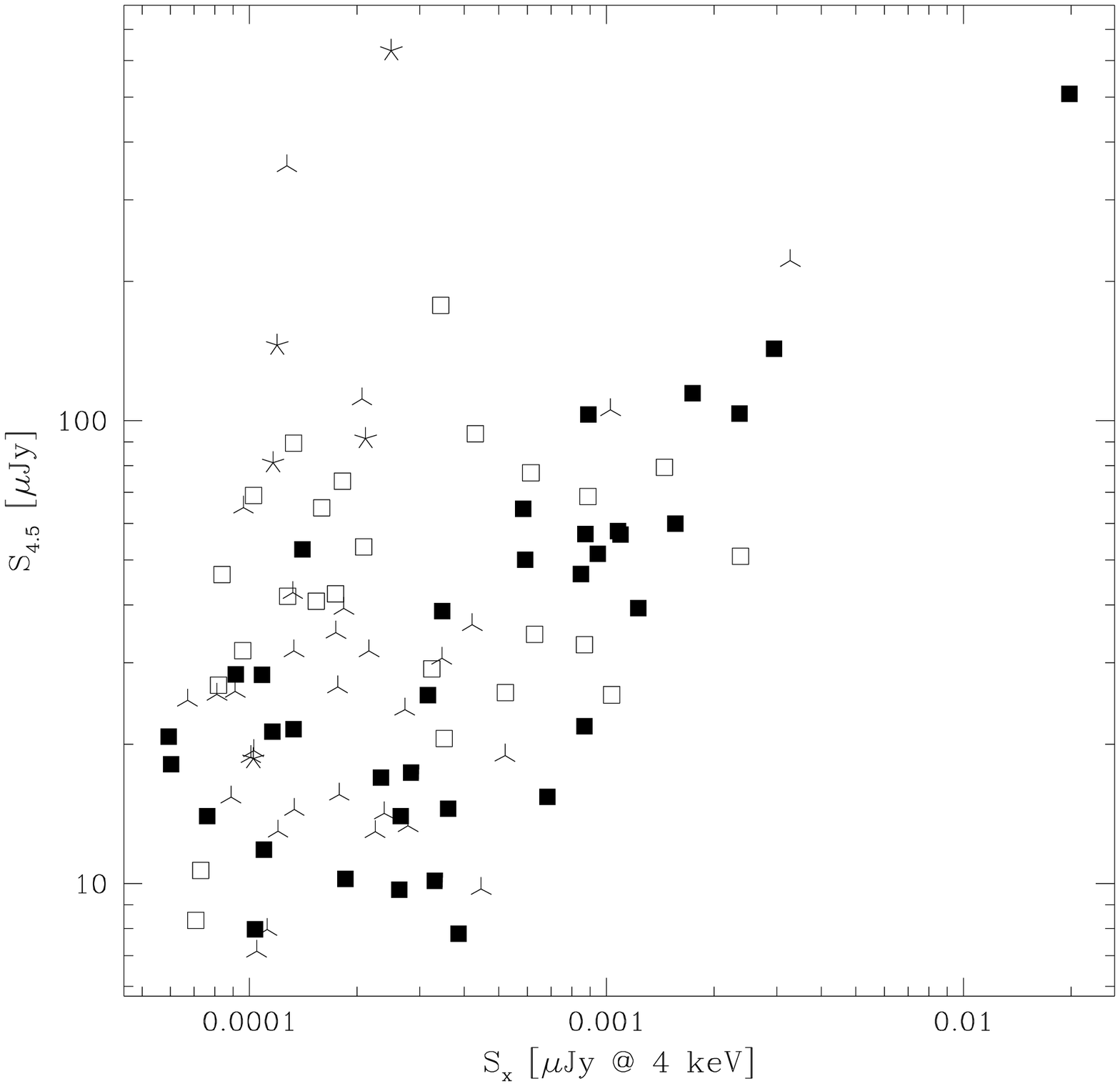,width=0.5\textwidth,clip=}\\
 \psfig{file=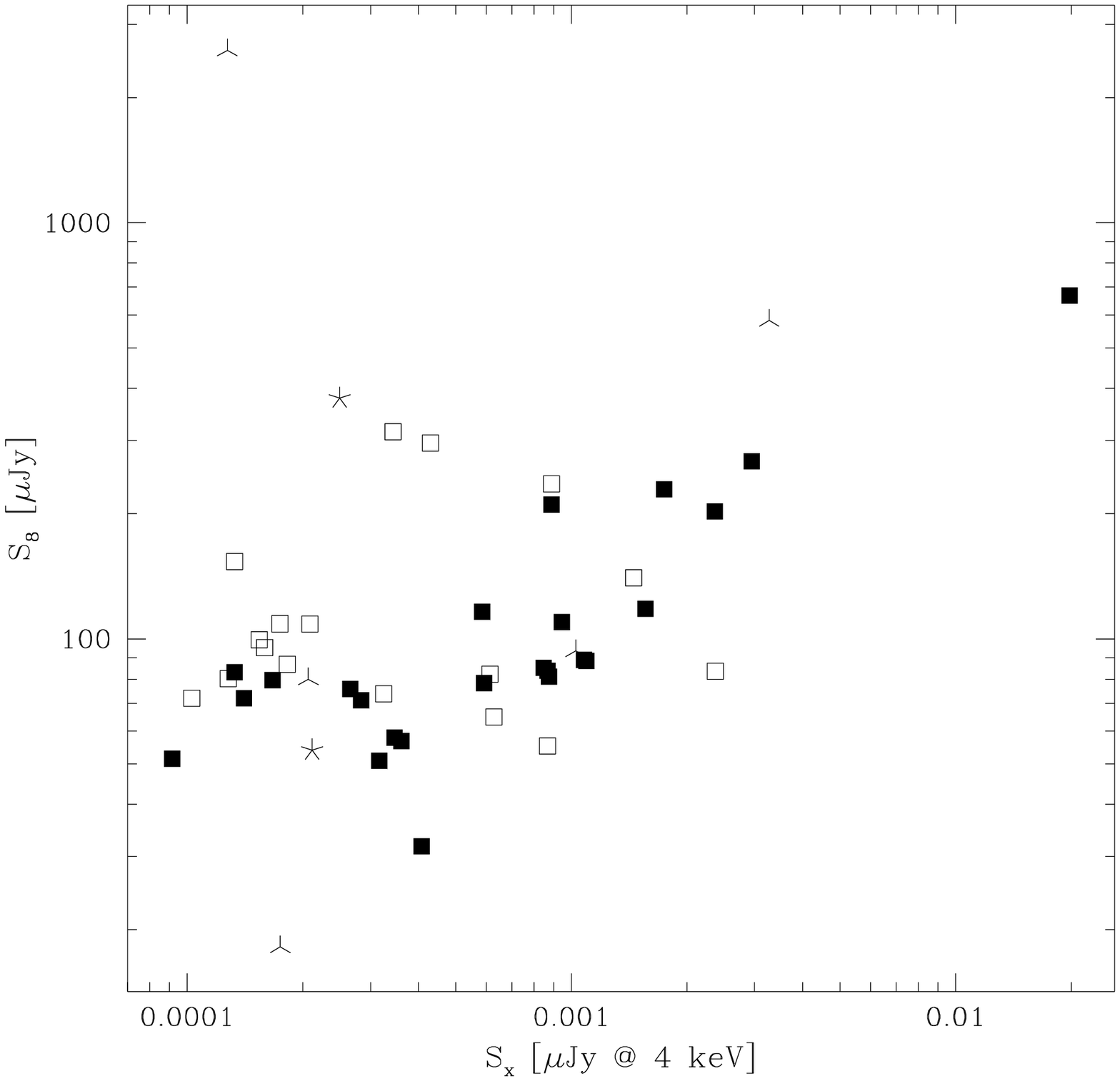 ,width=0.5\textwidth,clip=} &  \epsfig{file=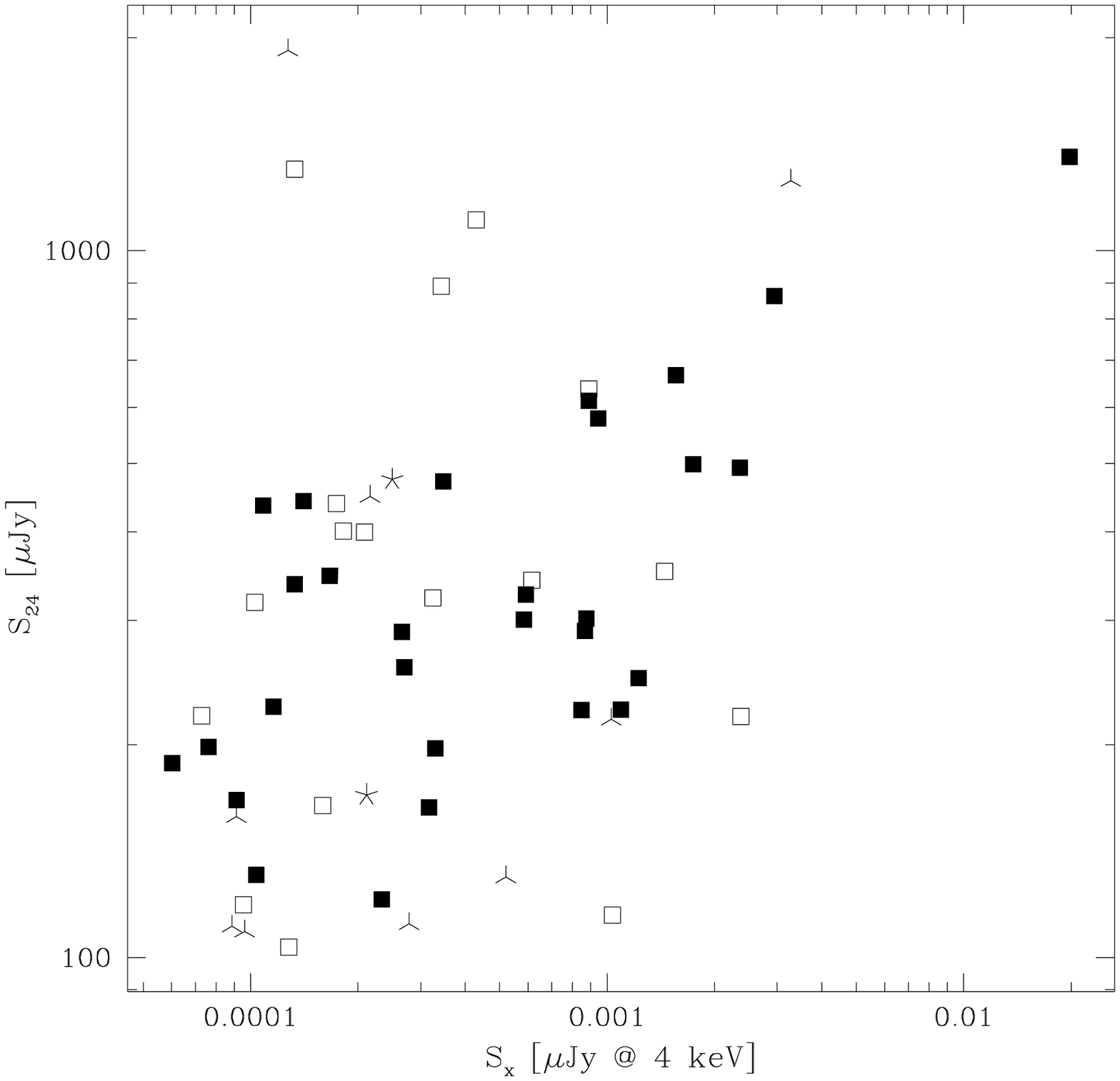,width=0.5\textwidth,clip=}\\
 \end{tabular}
\caption{Relation between the IRAC and MIPS 24 $\mu$m flux densities (in $\mu$Jy) and the X-ray flux densities at 4 keV (based on Chandra fluxes in the total 0.5-8 keV band).
In all panels filled squares are type-1 quasars; open squares are type-2 AGNs; 3-leg stars are spirals and starbursts; 5-leg stars early-type galaxies. Clockwise from top left to bottom right we have: correlation with the 3.6 $\mu$m, 4.5 $\mu$m, 8 $\mu$m IRAC and 24 $\mu$m MIPS fluxes.   
}
\label{corre}
\end{figure}

We have performed an X-ray to IR flux correlation study by exploiting our large sample and extensive IR coverage. We report in Fig. \ref{corre} plots of the X-ray flux densities at 4 keV (based on Chandra fluxes in the total 0.5-8 keV band) versus those in the IRAC and MIPS 24 $\mu$m bands for our various classes of sources. 

X-ray and mid-IR fluxes appear significantly correlated only for type-1 quasars and AGNs. The Spearman correlation coefficient for the 3.6 $\mu$m, 4.5 $\mu$m, 8 $\mu$m IRAC fluxes
and the MIPS 24 $\mu$m are C.C.=0.70, 0.69, 0.81 and 0.62, respectively: the tightest correlation is then with the 8 $\mu$m flux, while the one between the X-ray and far-IR 24 $\mu$m fluxes is less significant.

For the type-2 AGNs and for normal galaxies and starbursts (our sources in categories 3 and 4) the correlations are much poorer (Spearman C.C.=0.02, 0.16, 0.06, 0.2, in the same order; and C.C.=0.17, 0.16, 0.17, 0.40). Perhaps surprisingly, none of these results depend significantly on the X-ray bandwidth, neither the soft nor the hard X-ray one, adopted for the comparison.
Although not statistically significant, it is interesting to note that the scatter in the X-ray to IR correlation for type-1 AGNs, and even more for other sources, increases at the lower fluxes. This seems to be an intrinsic effect, not due to the increased photometric uncertainties at the lower fluxes, which are lower than the observed scatter.

This may support the case that, for type-1 AGNs, the IR and X-ray emissions originate from a single dominant physical process, i.e. energy production by gravitational accretion. Furthermore, absorption effects in the dust torus due to atomic photoelectric effects and dust extinction should be unimportant. 
This may be in keeping with the notions that: a) in type-1 AGNs the contribution of star-formation to the energy production is irrelevant compared with the QSO emission itself; b) the line-of-sight is not intersecting large column densities of absorbing material; c) the dust distribution covers only a fraction of the ionizing photons emitted by the central power source (see Sect. \ref{cove} about this point).

All this is evidently not true for the other classes of sources. As recently discussed in Lutz et al. (2004), the large contribution of star-formation, particularly to the mid- and far-IR emissions, is likely responsible for increasing the scatter going from type-1 to the type-2 AGN population. Eventually, the lack of apparent correlation between X-ray and IR emission in the X-ray subsample identified with normal or purely starburst galaxies in optical/IR shows that the two emissions may come from independent physical processes (quasar nuclear activity and stellar emission, either direct or dust-reprocessed). 

\begin{figure}
 \epsfig{file=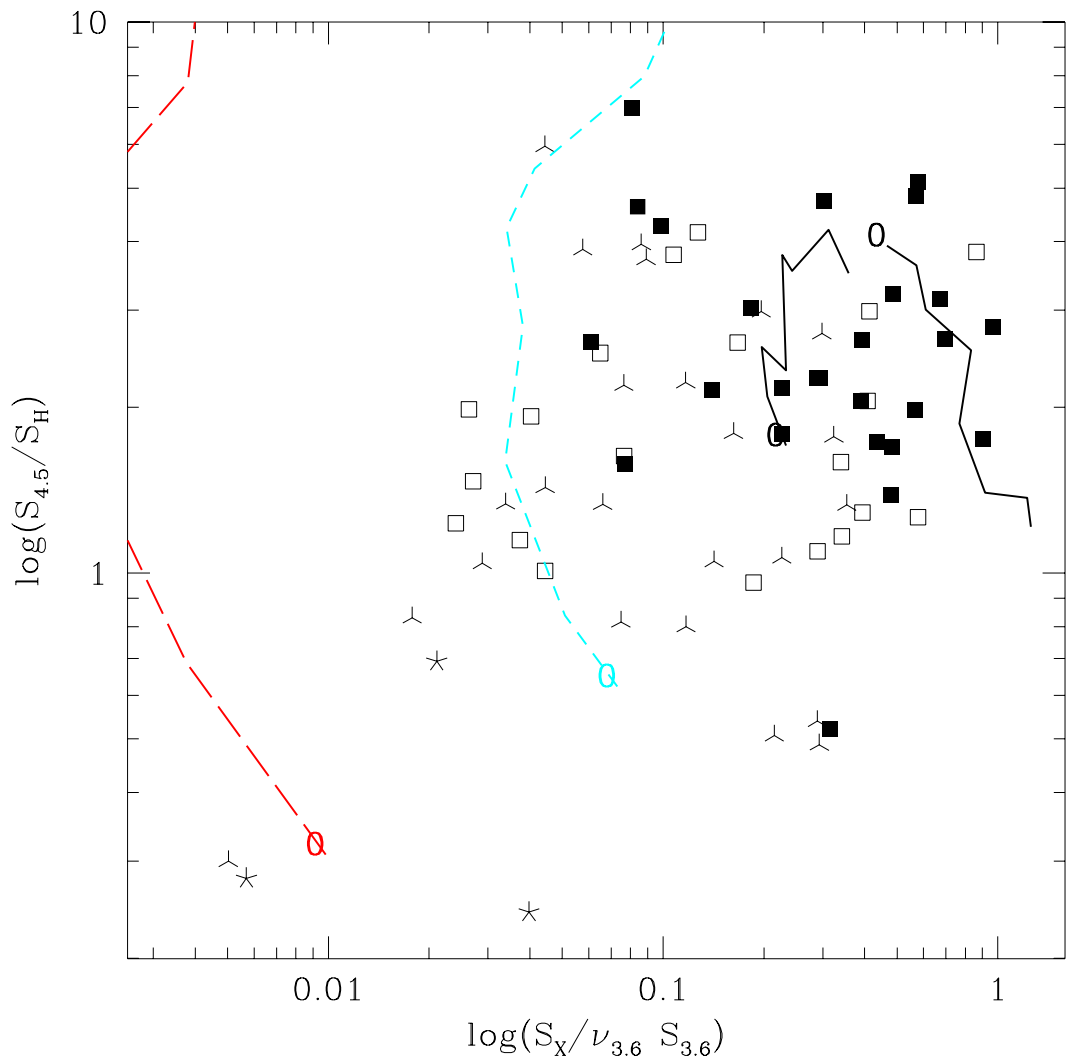,width=0.9\textwidth,clip=} 
\caption{IRAC 4.5 $\mu$m to H-band colors versus the ratios of the X-ray to IR fluxes. $S_X$ is the flux in the total 0.5 - 8 keV band, $\nu_{3.6}S_{3.6}$ is the IRAC channel-1 band flux.  Symbols as in Fig. \ref{diagno}.
The lines are the predicted broad-band colors as a function of z: the color for z=0 corresponds to the point marked with "0", and lines are drawn from here to points corresponding to z=0.5, 1, 1.5, 2, 2.5, 3 and 3.5. From right to left, the lines are:  type-1 quasar and Seyfert-1 (two continuous black lines), Seyfert-2 (cyan short-dashes), and Sb spiral/starburst (red long-dashes).
}
\label{X_36_17}
\end{figure}

Finally, an attempt to combine an optical-IR color criterion with the ratio of the IR to X-ray fluxes is reported in Fig. \ref{X_36_17}. Here the ratio of the X-ray to IR fluxes is compared with the IRAC channel 2 to H-band color. Type-1 quasars cluster around the upper right corner (high X-ray and mid-IR emissions), while type-2 AGNs and sources with starburst optical-IR colors scatter over a much wider region. 

In the same figure we report the predicted broad-band colors for spectral templates as a function of the source redshift. As in Fig. \ref{diagno}, for each lines the color for redshift z=0 corresponds to the point marked with "0", and lines are drawn from here to the points corresponding to z=0.5, 1, 1.5, 2, 2.5, 3 and 3.5. Going from the right to the left, the two continuous (black) lines are for a type-1 quasar and a Seyfert-1 templates, the (cyan) short-dash for a Seyfert-2, and the (red) long-dash for a Sb spiral/starburst.
The X to IR flux ratios of the templates come from observations of local sources (see Sects. \ref{3.1} and \ref{3.2}).
This confirms that for type-1 objects the evolution of flux ratios is very moderate, while it is more pronounced for type-2 objects and starbursts. 

An important point emerging from Fig. \ref{X_36_17} is that the X-ray to IRAC flux ratios $S_X/\nu_{3.6}S_{3.6}$ of our source sample are, almost without exception, inconsistent with an X-ray emission due to a starburst (whose z-dependent path in this color-color diagram should be close to the long-dashed line in the far left). This agrees well with the results of Fig. \ref{histo}.

\subsection{X-ray Hardness Ratio Analysis}

The spectral information contained in our deep Chandra X-ray survey provides additional value for source characterization. Given the limited depth and photon counts, the X-ray   spectral data were summarized in the usual form of a broad-band hardness ratio $HR$.  
This was defined as $ HR = \frac{(H - S)}{(H + S)} $, where H and S are the background-subtracted X-ray counts in the hard (2.0 - 8.0 keV) and soft (0.5 - 2.0 keV) bands, respectively.  Full details on the HR calculation cab be found in Manners et al. (2003).

\begin{figure}
 \epsfig{file=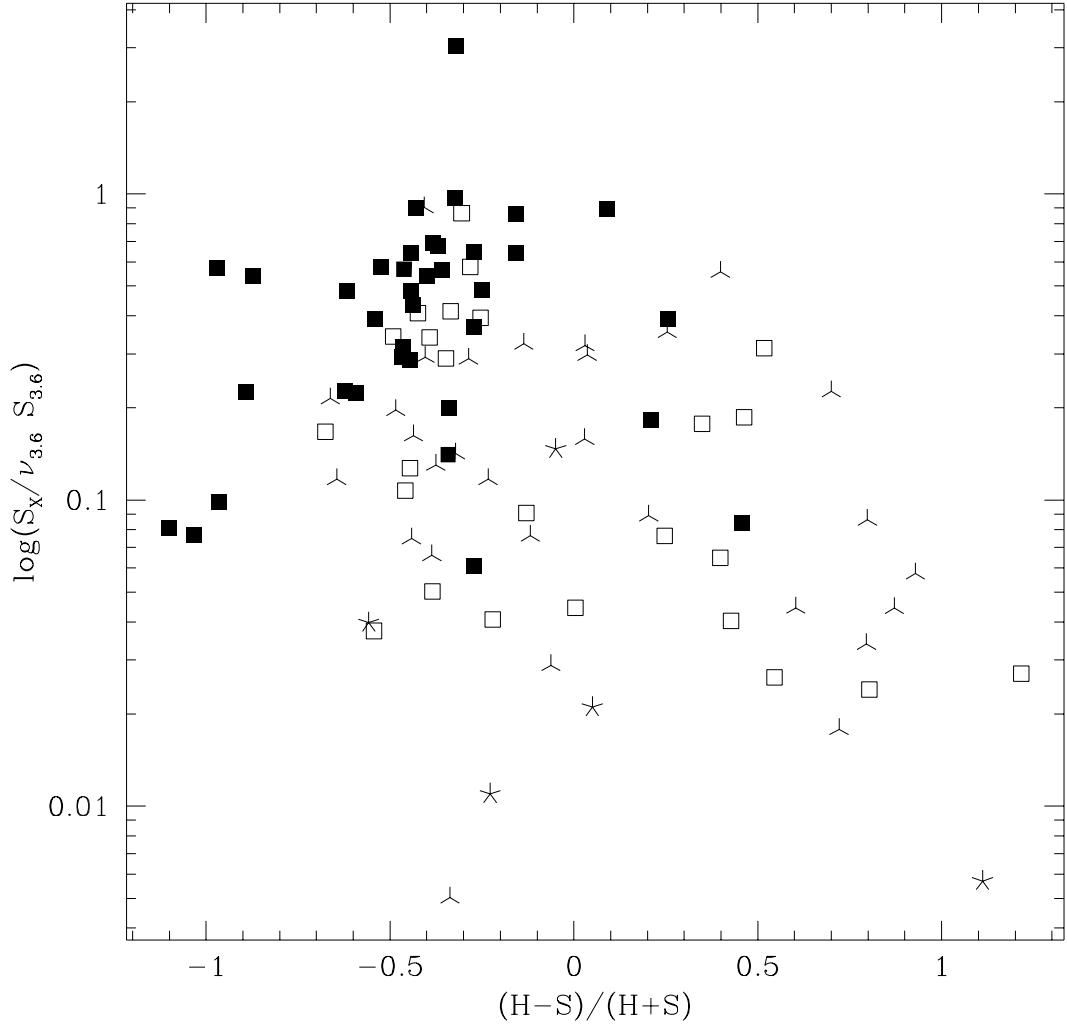,width=0.9\textwidth,clip=} 
\caption{The X-ray hardness ratio versus X-ray to infrared flux ratio. $ HR $ is defined as $\frac{(H-S)}{(H+S)}$, where H and S are the net X-ray counts in the hard (2.0 - 8.0 keV) and soft (0.5 - 2.0 keV) bands. $S_X$ is the usual total 0.5 - 8 keV flux.}
\label{hr}
\end{figure}

Figure \ref{hr} plots such hardness ratio against the X-ray to IR flux ratio. A first obvious effect is the clustering of type-1 quasars and AGNs around values of $HR\simeq -0.5$ and values of the X-ray to IR flux ratio from 0.3 to 1. Such values of $HR$ corresponds to the classical $\alpha_X=0.7$ unabsorbed spectrum. 

Type-2 AGNs tend to spread out of the canonical AGN-1 region and to display significantly harder X-ray spectra and a lower X-ray emissivity. Both these effects are easily interpreted as due to X-ray photoelectric absorption. A similar depression of the X-ray emissivity is apparent in Fig. \ref{corre} correlating the IRAC channel-1 and total X-ray fluxes. The population of optically-classified normal galaxies and starbursts display even further spread out of the type-1 region.

What is interesting to note about this and Fig. \ref{X_36_17}, however, is that hardening of the X-ray spectrum and decrease of the mid-IR AGN excess do not correspond to a dramatic decrease of the source X-ray emissivity, as might be expected in the presence of strong absorption in either X-rays or the IR, or due to a change in the source population: with few exceptions, the values of the X-ray to IR flux ratio span a range not larger than a factor of thirty altogether.

\subsection{Redshift and Luminosity Distributions}\label{red}

We now undertake the more uncertain task of exploiting the information on cosmic distances for our combined Chandra/SWIRE sample, mostly based on the photometric redshift technique. Until we acquire a more systematic spectroscopic coverage for this sample, such information on redshifts should be taken in a statistical sense rather than to value for individual sources. This is particularly the case for the featureless type-1 AGNs.

We report in Fig. \ref{z} the photometric redshift distributions for our sample sources split into three different classes: type-1 AGNs (shaded histogram), type-2 AGNs (dotted histogram), and normal galaxies and starbursts (our previous classes 3 \& 4, dashes).
The total distribution is reported as a continuous-line. It is evident that our source sampling mainly detects galaxies and AGN between $z\sim 0.5$ and $z\sim 1.5$, with a maximum at $z=1$ and a fairly sustained tail up to $z=2.5$.

This distribution is interestingly similar to that reported by Zheng et al. (2004), which  comes from an optical spectroscopic survey of the deep Chandra survey in the CDFS, complemented with a photometric redshift analysis similar to ours. In CDFS the observed z-distribution shows a narrow peak at $z=0.6$, which we do not observe, probably due to 
a large clustering overdensity in their sample.
Compared to Zheng et al., a substantial advantage of our analysis is in the very extensive coverage of the mid- and far-IR bands which contain important diagnostic information and brings improved capabilities for redshift determination.

It is important to note that our sample is very representative of the sources of the XRB, if we consider that it includes 84\% of the Chandra X-ray sample complete to $S_{0.5-8 keV}\simeq 2\ 10^{-15}\ erg/cm^2/s$: sources detected at this limit contribute roughly 60\% of the XRB (Manners et al. 2003; see also Hasinger et al. 2001; Brandt et al. 2001; Giacconi et al. 2001).
Our present results confirm earlier reported evidence (Hasinger 2003; Szokoly et al. 2004; Gilli et al. 2003) that most of the X-ray sources of the XRB are at redshift lower than 1.5.

As for the individual population contributions, type-1 AGNs are spread over a rather large z-interval to $z>2$, while type-2 objects and starburst/normal galaxies appear more concentrated at $z=0.5$ to 1.5. Such evidence of a moderately low median redshift for the sources of the XRB, which we confirm here in a statistical sense, demands significant revision of the XRB synthesis models (Franceschini, Braito \& Fadda 2002; Gandhi \& Fabian 2003).

We report in Fig. \ref{Lz} the population of the Chandra/SWIRE sources in the X-ray luminosity - redshift plane. 
L$_{0.5 - 8 keV}$ is the rest frame X-ray luminosity $L_X=4\pi d_L^2 S_X/(1+z)^{1-\alpha_X}$, with $d_L$ being the usual luminosity distance computed for the $H_0=71$, $\Omega_m=0.27$, $\Omega_\Lambda=0.73$ case. The spectral slope ($\alpha_X$) used to derive the rest frame luminosities was calculated based on the X-ray hardness ratio (HR) for each source: 
\begin{equation}
\alpha_X = \frac{\log \left[(1-HR)/(1+HR)\right]}{2 \log 2}
\end{equation}
The flux $S_X$ is calculated from the observed X-ray count-rate adopting the $\alpha_X$ spectral slope, in a consistent way.

After the analysis of Fadda et al. (2002) and Franceschini et al. (2003), the X-ray luminosity is by itself an important discriminant of the primary power sources. On statistical grounds, and also supported by synthetic modelling of the X-ray emission by young stellar populations (Persic et al. 2004), it turns out that an X-ray luminosity of $L_{0.5-8 keV} \simeq 3\times 10^{42}\ erg/s$ ($L_{2-10 keV}\simeq 1.5\times 10^{42}\ erg/s$) is the maximum achievable by an ultra-luminous starburst. This X-ray luminosity corresponds to a rate of star formation of $\sim 1500\ M_\odot/yr$ assuming a Salpeter stellar IMF between 0.1 and 100 $M_\odot$, again a limit achievable by a starburst in the most luminous ULIRGs.

An important conclusion then emerges from the comparison of this limit with the X-ray 
luminosities estimated for our sources: if we exclude a few among the most local, typically at $z< 0.5$, the vast majority exceed this X-ray luminosity value. Our conclusion is that, with few exceptions, our sources are dominated by quasar emission.

A further diagnostic is provided by the ratio of the X-ray to bolometric luminosities $L_X/L_{bol}$, where $L_{bol}$ is determined from our combined fits of the near-, mid- and far-IR SEDs.
We report in Fig. \ref{LxLfir} this $L_X/L_{bol}$ luminosity ratio against the total 0.5 - 8 keV band X-ray luminosity $L_X$. Our sources display a fairly large range of $L_X/L_{bol}$ values, with the X-ray luminosity ranging from $\sim 0.5\%$ to somewhat more than 10\% than $L_{bol}$. For comparison, in samples of local ULIRGs, Braito (2003) and Franceschini et al. (2003) find that a discriminant between a starburst-dominated and quasar-dominated ultraluminous source may be set at $L_X/L_{bol}\simeq 2\ L_{2-10}/L_{bol} \simeq 10^{-4}$. Indeed, in these studies, local ULIRGs with evidence of AGN activity in the IR, like Mkn 231 and the Superantennae, have typical values of the X-ray luminosity around 0.1\% of the bolometric ($L_X/L_{bol}\sim 10^{-3}$).   The X-ray prominence of our sample sources is evidently the result of our primary X-ray selection. 

Based on the results of this Sect. and on Figs. \ref{histo} and \ref{X_36_17}, we confirm that, with almost no exception, the combined SWIRE-Chandra sources are dominated in X-rays by AGN emission. A dominant fraction of the XRB itself is then proven to come from AGNs. 

For more than one third of these sources ($>38\%$, our classes 3\&4 plus Chandra objects undetected in the IR), our analysis of the optical-IR SEDs did not reveal AGN signatures, although they are AGN-dominated in X-rays. In some cases this lack of evidence may be due to the moderate depth of our survey in the IRAC channels 3 and 4, preventing in half of the sample a proper characterization of the SED at wavelengths (5.8 and 8 $\mu$m) where we would expect the AGN torus emission to emerge. 
In at least 14 sources, however, the observational SEDs have good IRAC coverage and are indeed best-fit by a starburst/normal galaxy template.
In such cases, the only way we can interprete the elusiveness of the AGN in the IR is by assuming a very large dust column density implying self-asorption in the near-IR torus emission.

\subsection{Statistical Analysis of IR and X-ray selected SWIRE AGNs }\label{stat}

In an attempt to understand the completeness of our AGN selection, we compare in Fig. \ref{other} the optical-IR colors of our SWIRE-Chandra AGNs with those of the SWIRE sources without Chandra counterparts. In the top panel of the figure we exploit the i-band to IRAC channel 4 color versus the g-i flux ratio, in the bottom panel we use the i-band to IRAC channel 3. The large symbols correspond to our Chandra AGNs, whereas the small (red) circles correspond to the SWIRE sources undetected by Chandra. It is evident from the figure that a region of the color-color plots (in the upper left corner) identifies in a very efficient way the Chandra type-1 AGNs, which are very blue in the optical and very red in the IR. 
On the contrary, the majority of the type-2 AGNs fall outside this region.

The Chandra-undetected sources tend to occupy a complementary region of the color-color plane of that filled by type-1 AGNs. Only a small fraction of Chandra-undetected sources (7 in the top and 7 in the bottom panels) fall inside it. Then the first important conclusion is that our SWIRE-Chandra selection seems rather efficient in identifying the bulk of the type-1 AGN population in the area: if anything, the Chandra-undetected sources (making up 90\% of the SWIRE population) may hide no more than an additional 20-30\% of the Chandra type-1 AGN population.

The second point emerging from Fig. \ref{other}, however, is that a $(S_I/S_{8})$ versus $(S_g/S_I)$ color classification hardly distinguishes the type-2 AGNs: many of the Chandra types-2 are spread in a region of the plot occupied by X-ray silent normal and starburst galaxies
(we cannot use for the present analysis the optimal $S_H/S_{4.5}$ versus $S_H/S_R$ criterion in Fig. \ref{diagno}, because we do not have a full-map coverage in the H-band).
In conclusion, we cannot exclude at the present stage that some fraction of the SWIRE population undetected by Chandra hides obscured AGNs.

Assuming that we have detected with Chandra the bulk of the AGN population in the area, we can derive constraints on the nature of the Spitzer-selected sources from a comparison with the X-ray map. We are particularly interested in the sample of MIPS 24 $\mu$m sources, since these are expected to contribute substantially to the CIRB background. From Papovich et al. (2004) the MIPS source contribution to the CIRB 
ranges from 40\% at $S_{24} > 0.4\ mJy$ to 80\% at $S_{24}>60\ \mu Jy$. The number counts of SWIRE sources in our 17'x17' Chandra field are reported in Fig. \ref{fig1}. 

We find that above $S_{24} > 0.4\ mJy$ there are 126 MIPS sources, 18 of which are Chandra AGNs (14\%). Above 300 $\mu$Jy there are 211 MIPS sources in the Chandra field, 27 of which are Chandra detections (13\%). The fraction of X-ray counterparts decreases steadily at fainter IR fluxes, reaching 9\% at 200 $\mu Jy$ (39 Chandra out of 433 MIPS sources) and 6\% at 100 $\mu Jy$ (59 out of 988). Considering the X-ray to 24 $\mu$m flux correlation (Fig. \ref{corre}), however, this last figure turns out to be already affected by incompleteness due to the X-ray flux limit. These numbers have in any case to be increased by a factor 1.2-1.3 to account for the type-1 Chandra-undetected objects (see Fig. \ref{other}).

The above numbers appear quite consistent with the AGN fraction for faint 15 $\mu$m sources 
estimated by Fadda et al. (2002) at comparable flux limits ($>0.4$ mJy, for which a 15 $\pm$5\% AGN fraction was estimated). Our improved statistics and sensitivity below this limit shows that the AGN fraction does not seem to increase at fainter fluxes. It remains to be seen, however, if an additional population of X-ray weak type-2 AGNs are hidden among SWIRE sources with normal (starburst) optical-IR colors. While an indirect argument against this is mentioned in Sect. \ref{con}, quantifying it will require either much deeper X-ray imaging or, better, deep imaging in very hard ($>10$ keV) X-rays.

\subsection{Estimating the Dust Covering Fraction in AGNs }\label{cove}

To constrain the dust covering fraction in our type-1 AGN sample, we compare in Fig. \ref{cov} an estimate of the AGN torus luminosity due to dust re-radiated emission, and directly observed by Spitzer, with the AGN bolometric luminosity as inferred from the X-rays. 
We have estimated the former through the observed source luminosity in the MIPS
24$\mu$m and IRAC 8$\mu$m channels: for type-1 AGNs, these roughly correspond to the expected emission peak by the AGN dust torus for our typical source redshifts (the dust torus emission is expected to peak at $\lambda \sim 5-10 \mu$m, as observationally seen e.g. by Elvis et al. 1994, and expected from theoretical modelling, Efstathiou \& Rowan-Robinson 1995; Granato et al. 1997).
The basic assumption of our analysis is that the fluxes in the 8$\mu$m and 24$\mu$m channels for our type-1 AGNs are dominated by the AGN torus emission rather than by ISM-reprocessed stellar light, which is supported by the power-law shapes of their IR SEDs
shown in Figs. \ref{figSED1} to \ref{figSED10}, very inconsistent with those of normal starburst galaxies.

These AGN IR luminosities are compared with the bolometric ones, $L_{bol,X}$, that we estimated from our observed 0.5-8 keV flux of the Chandra sources by applying a bolometric correction based on Elvis et al. (1994). To obtain this correction factor we have integrated their average quasar SED and found a value of $\simeq 20$ to go from $L_X (0.5-8 keV)$ to bolometric.

In the top panel of Fig. \ref{cov} the bolometric emission based on the X-ray flux is compared with the AGN IR emission $L_{AGN,IR}$ estimated from the 8$\mu$m flux as 
$L_{AGN,IR}=ln(10)\times \nu L(\nu)|_{8\mu}$ (corresponding to the energy in one decade in wavelength centered at 8$\mu$m). In the lower panel the comparison is made with the 24$\mu$m-based estimate: $L_{AGN,IR}=ln(10)\times \nu L(\nu)|_{24\mu}$.  

The bolometric AGN luminosities inferred from the X-ray flux and the AGN IR luminosities turn out to be fairly well correlated in Fig. \ref{cov}a, while the correlation is poorer in \ref{cov}b (the 24$\mu$m, see also Fig. \ref{corre}). The continuous and dotted lines in the figure correspond to the predictions in which 100\% and 10\% of the AGN bolometric emission are reprocessed by dust into the IR, assuming that both $L_{AGN,IR}$ and $L_{bol,X}$ are dominated by AGN emission.

Under the same hypothesis of an AGN dominance in the X-ray and 24-8 $\mu$m fluxes (argued on the basis of the power-law IR spectra for type-1), the position of each sources compared with these two lines can be considered to bear an indication of the covering factor of the circumnuclear dust.
This covering factor ranges between values of 10 and 100\%. Occasionally, 
$L_{AGN,IR}$ exceeds $L_{bol,X}$, particularly when the former is computed from the 24$\mu$m flux (panel [b]), which may indicate a possible starburst contribution to the IR emission in some of these sources.

\section{DISCUSSION AND CONCLUSIONS }\label{con}

A clear limitation in our analysis may be seen in the shallow depth achieved with the 
IRAC channel-3 and 4, which would otherwise be important for source diagnostics. In many 
instances, even the upper limits of our IRAC survey have set useful constraints on the spectral fits. Deeper IRAC surveys will soon offer such improved capabilities.

We have found that 39\% of the Chandra sources are dominated by type-1 
AGN emission (QSOs or Seyfert 1), 23\% are type-2 AGNs, while the remaining 38\% fraction 
are consistent with starburst-like or even normal galaxy spectra in the optical/IR. 

\begin{figure}
 \epsfig{file=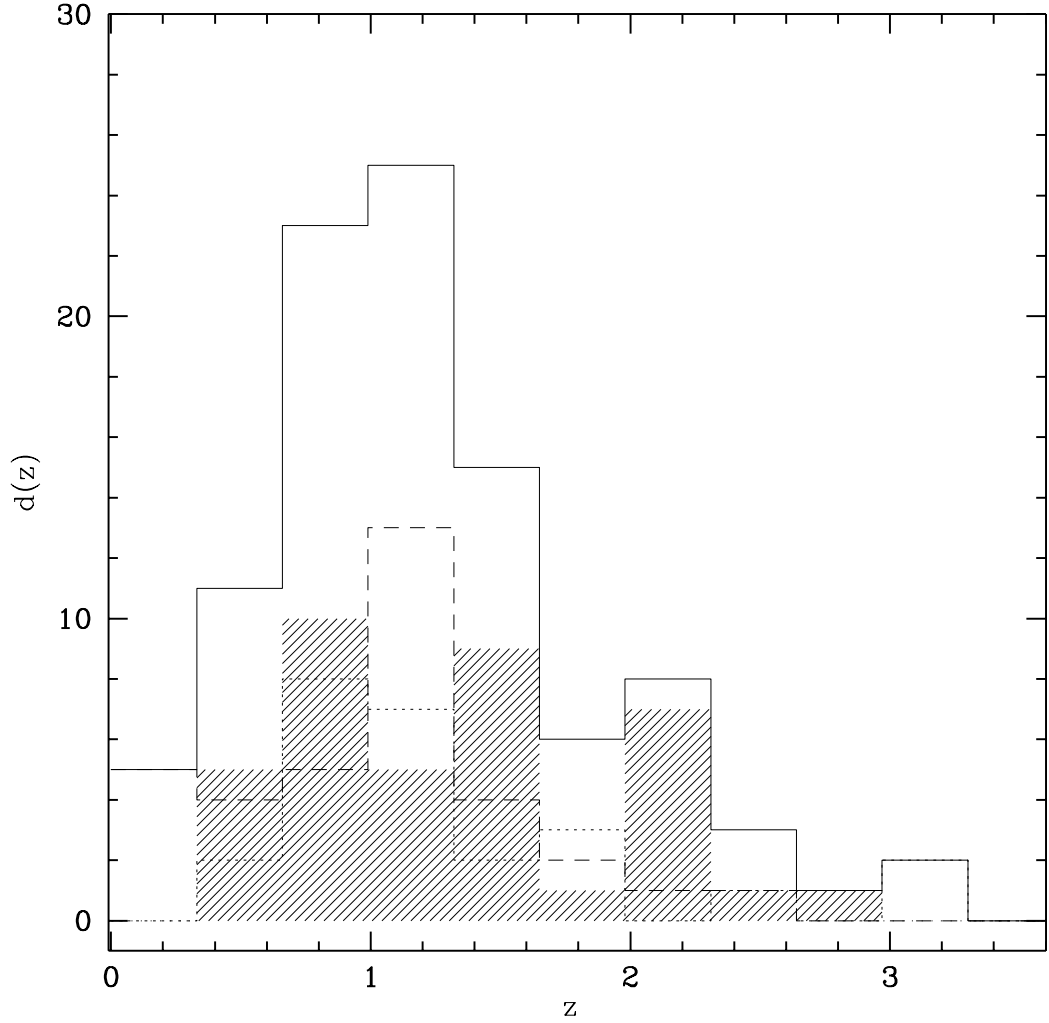,width=0.9\textwidth,clip=} 
\caption{Redshift distributions for 99 sources detected by Chandra in ELAIS-N1 and identified with galaxies or AGNs (3 stars are excluded). Three classes of sources are shown: type-1 AGNs (shaded histogram), type-2 AGNs (dotted histogram), and normal galaxies and starbursts (our previous classes 3 \& 4, dashes).  Symbols as in Fig. \ref{diagno}.
}
\label{z}
\end{figure}

\begin{figure}
 \epsfig{file=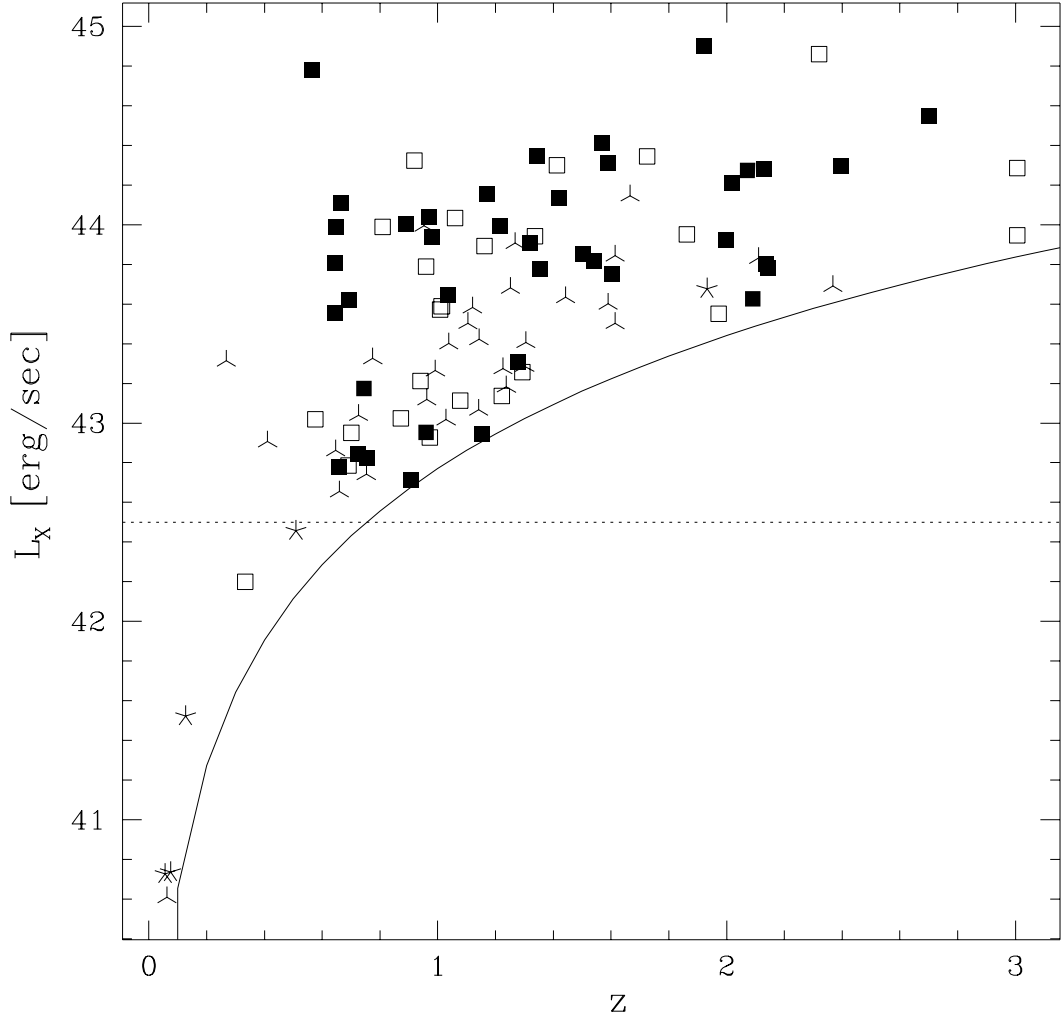,width=0.9\textwidth,clip=} 
\caption{Distribution of rest frame 0.5-8 keV X-ray luminosity and redshift for 99 sources detected by Chandra in ELAIS-N1. The bulk of redshifts are photometric, and are to be taken as just indicative for the type-1 AGNs. Source symbols as in previous figures. The dotted line marks the upper boundary of the X-ray luminosity produced by an ultraluminous starburst. 
The continous line shows the $S_X=10^{-15}\ erg/cm^2/s$ observational flux limit. 
}
\label{Lz}
\end{figure}

\begin{figure}
 \epsfig{file=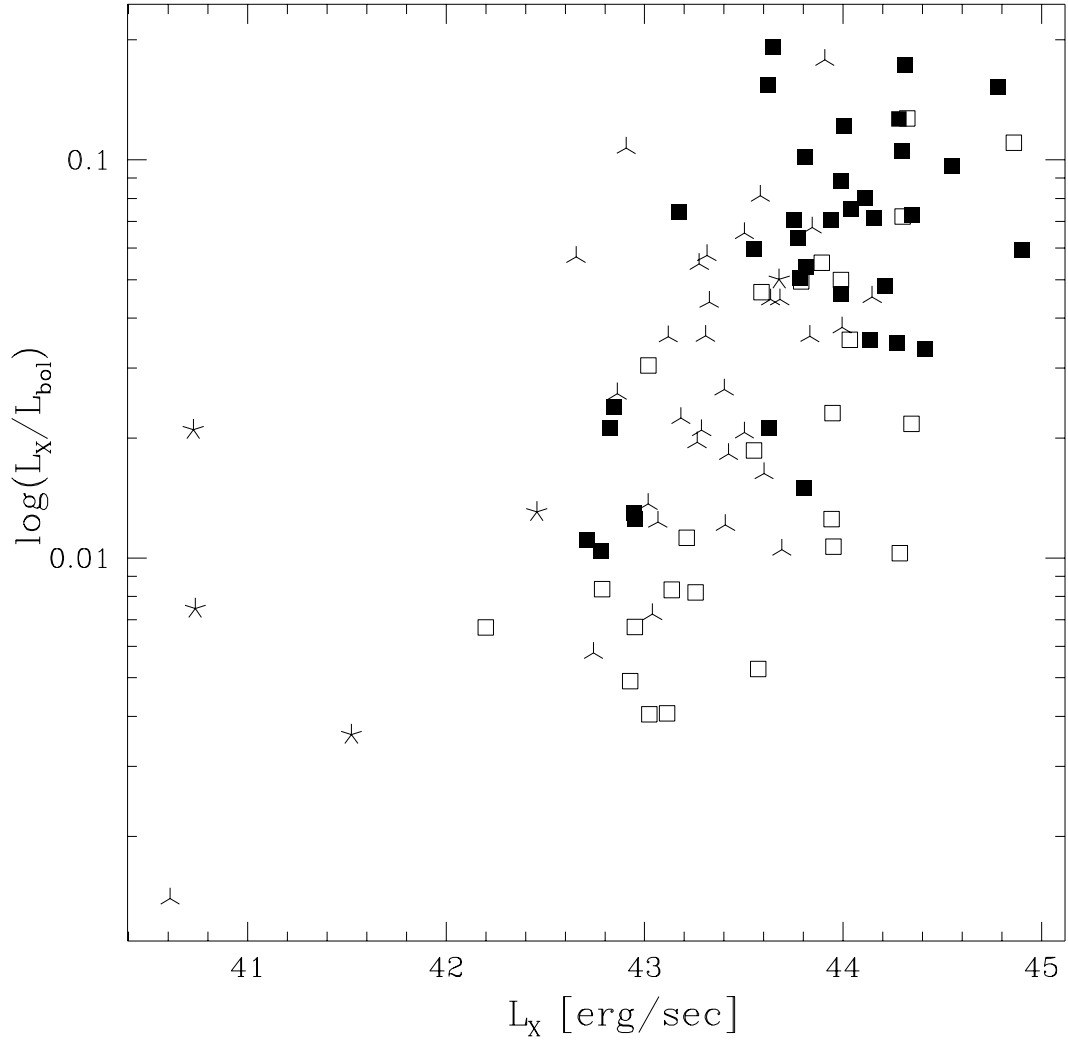,width=0.9\textwidth,clip=} 
\caption{Plot of the ratio of the total 0.5 - 8 keV X-ray to bolometric luminosity as a function of $L_X$.  See also captions to previous figures.
}
\label{LxLfir}
\end{figure}

\begin{figure}
 \epsfig{file=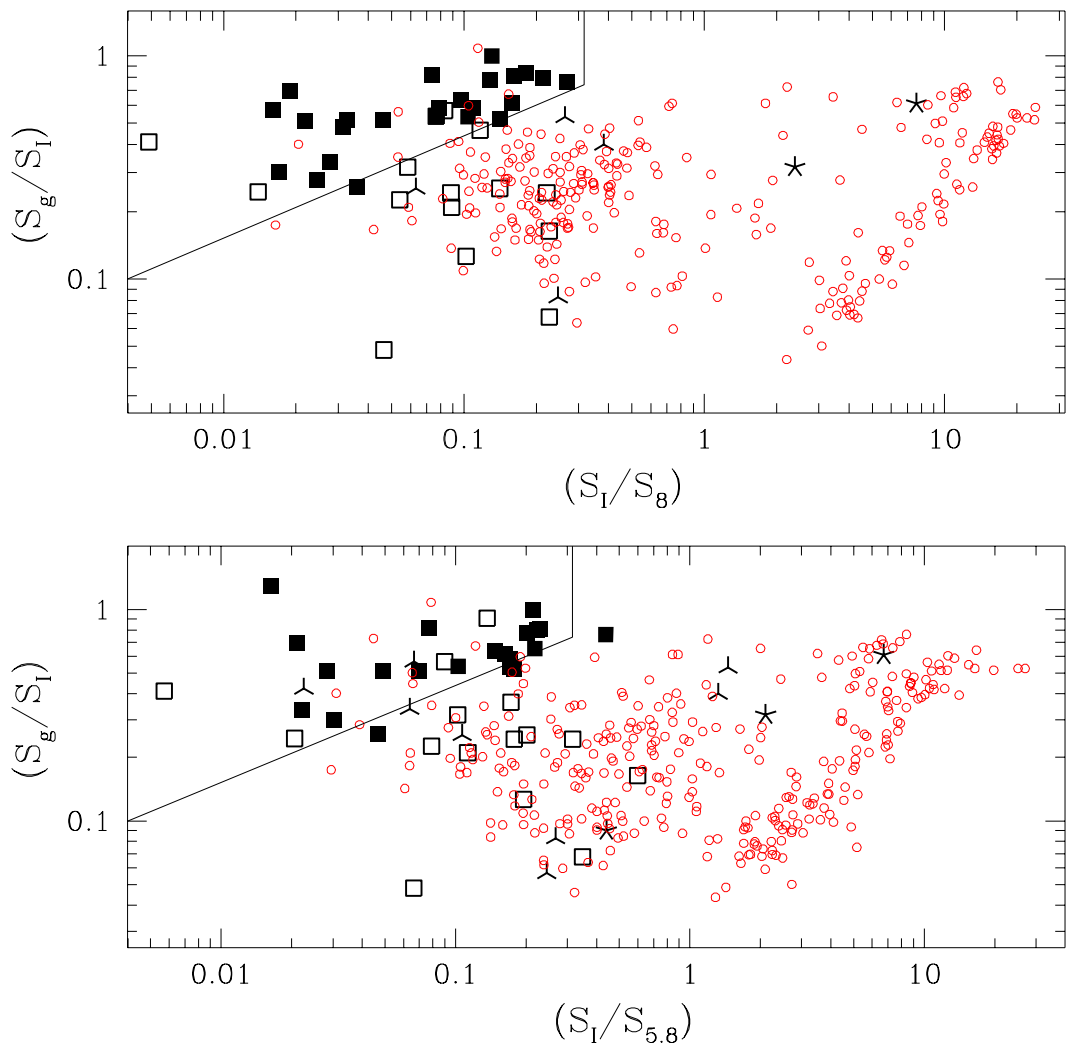,width=0.9\textwidth,height=0.9\textwidth,clip=} 
\caption{Comparison of the optical-IR colors of SWIRE-Chandra AGNs (large symbols, see caption to Fig. \ref{diagno}) with those of the SWIRE sources without Chandra counterparts (small red circles). Top panel exploits the i-band to IRAC channel 4 color versus g-i, the bottom panel uses the i-band to IRAC channel 3. The region in the upper left corner, delimited by the following vertices (-2.4; -1), (-.5; -0.13) and (-.5; 1),
contains virtually all the Chandra type-1 AGNs in both panels. The majority of the type-2 AGNs and of the Chandra-undetected galaxies fall outside this region.
}
\label{other}
\end{figure}

\begin{figure}
 \begin{tabular}{c}
 \epsfig{file=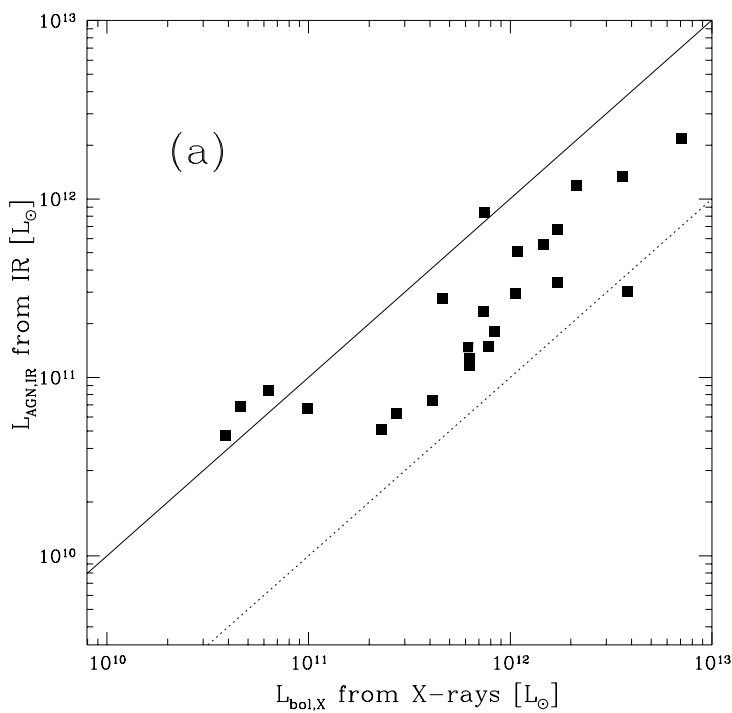,width=0.6\textwidth,height=0.6\textwidth,clip=}      \\
 \epsfig{file=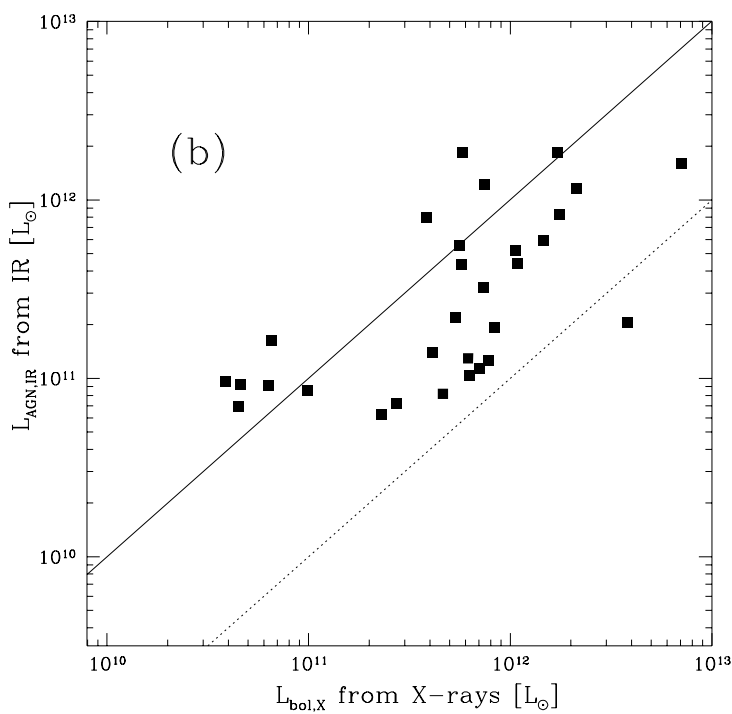,width=0.6\textwidth,height=0.6\textwidth,clip=}      \\
 \end{tabular}
\caption{The luminosity of the AGN dust emission component in type-1 estimated from the Spitzer IR luminosity versus the bolometric luminosity derived from the X-ray flux. 
In panel (a) the dust emission is estimated from the IRAC 8$\mu$m, in panel (b) from the MIPS 24$\mu$m luminosities.
The continuous line marks a covering fraction of 100\% (i.e. $L_{AGN,IR}/L_{bol,X} = 1$) and the dotted line is for covering fraction of 10\%.  Only sources detected at 8 and 24 $\mu$m are reported in the respective panels.
}
\label{cov}
\end{figure}

Based on our estimated X-ray luminosities and the X-ray to IR flux ratios, we have shown that, with few exceptions, all the Chandra sources in our sample (including those with galaxy-like optical/IR spectra) are dominated in X-rays by AGN emission. 
This allows us to estimate the global fraction of type-1 (our class-1) to type-2 AGNs (our classes 2, 3 and 4): assuming conservatively that all Chandra sources undetected by SWIRE are type-2 AGNs, the fraction of type-1 would still be half at most of the type-2 objects, and the type-1 over total AGN fraction not less than one third (39 objects out of 119). 
This type-1 AGN fraction is definitely larger than the canonical value 1/4 to 1/5, or even 1/10, needed to explain the XRB under the strict hypotheses of the standard XRB models (e.g. Comastri et al. 1995; Gilli et al. 2001). 
A solution is likely to come from Fig. \ref{z}, showing that an important fraction of the sources of the XRB are found around z=1 (and 70\% at $z < 1.5$): the XRB comes mostly from type-2 AGNs at moderate to low-z, much lower than previously inferred from type-1 quasar surveys (typically found at $z \sim 2$).

Our analysis of the mid-IR MIPS 24 $\mu$m selected sources, composing much of the CIRB background, has shown that the fraction of those sources dominated by an AGN (either type-1 or type-2) is around $\sim 10\%$ to 20\% down to 0.3 mJy, and may decrease at fainter fluxes. This is consistent with previous findings (Fadda et al. 2002). 
However, this information does not translates directly into a constraint on the AGN contribution to the CIRB background, which is at the moment rather uncertain. 

Indeed, on one side the above AGN fraction is to be considered as an upper limit to the bolometric AGN contribution to the CIRB, because the AGN IR emission peaks at much shorter wavelengths (rest-frame 10 $\mu$m) than the CIRB. On the other side, the available data, like those discussed in the present paper, cannot rule out yet the existence of an additional AGN population of Compton-thick AGNs undetectable in X-rays that could contribute to the CIRB background beyond our current $10\%$ limit (e.g. Farrah et al, 2002; Wilman et al. 2003). 

It may be worth to note that 
the locally observed mass density of supermassive black holes (BH) in galaxies may set a constraint here, although indirect and somehow model-dependent. Putting together 
the optical/X-ray type-1 quasar counts and the XRB spectral intensity, Fabian (2003) (see also Franceschini 2004) 
argues that, assuming a standard 10\% efficiency of gas accretion in AGNs, the whole BH mass density is explained by normal Compton-thin X-ray/optical AGNs. This sets an upper limit of 20\% at most to the local BH density possibly coming from Compton-thick accretion, which implies a modest addition, if any, to the above numerology and a minor contribution altogether by Compton-thick AGNs to the CIRB. 

To summarize, we have exploited combined Chandra and Spitzer multi-wavelength imaging and optical/near-IR data to perform a systematic analysis of faint far-IR and X-ray 
selected sources. Our observations substantiate the concept that the two well separated observational windows tend to select cosmic sources powered by fundamentally different processes, the far-IR probing preferentially stellar thermonuclear reactions, while the X-ray samples AGN accretion. Our results show that, on one hand, the IR observations would be capable of detecting with high efficiency (even at the moderate depths of the SWIRE Legacy survey) the XRB sources. On the other hand, we have shown that for a significant fraction (from 15 up to 40\%) of the X-ray/IR sources in common only deep X-ray observations have revealed the presence of moderately luminous and strongly absorbed AGNs.


\acknowledgments
Support for this work, part of the Spitzer Space Telescope Legacy Science 
Program, was provided by NASA through an award issued by the Jet Propulsion 
Laboratory, California Institute of Technology under NASA contract 1407.
This work was also supported by the European Community RTN Network "POE" (grant nr. HPRN-CT-2000-00138).



\clearpage

\clearpage

\clearpage

\clearpage


\clearpage

\clearpage

\begin{deluxetable}{lcccc}
\tablewidth{0pt} 
\tablecaption{Identification Statistics of Spitzer and Chandra Sources}
\tablehead{                              
\colhead{SWIRE channel} & \colhead{SWIRE sources} & $S_{lim}$ & \colhead{Chandra Detections} & \colhead{Percentage}  \\
}
\startdata
IRAC 1 &     3342 sources & 6 $\mu Jy$   & 100   &  3.0\%  \\
IRAC 2 & 3306 sources     & 6 $\mu Jy$   & 100   &  3.0\%  \\
IRAC 3 & 559 sources      & 25 $\mu Jy$  & 56    & 10.0\%  \\
IRAC 4 & 432 sources      & 32 $\mu Jy$  & 51    & 11.8\%  \\
MIPS24 & 988 sources      & 100 $\mu Jy$ & 59    &  6.0\%  \\
\enddata
\end{deluxetable}

\begin{deluxetable}{lll}
\tablewidth{0pt}
\tablecaption{Details of the spectral fits used with `Hyperz'.}
\tablehead{
\colhead{Templates} & \colhead{Spectral type} & \colhead{Reference}  \\
}
\startdata
 Ell(1.5)   & Ell (t = 1.5 Gyr)&  S98 (Silva {\it et al.} 1998)\\
 Ell(2)     & Ell (t = 2 Gyr)&  S98\\
 Ell(4)     & Ell (t = 4 Gyr)&  S98\\
 Ell(13)    & Ell (t = 13 Gyr)&  S98\\
 Sa         & Sa            &  S98, PAH modified using ISO PHT-S spectra\\
 Sb         & Sb            &  S98, PAH modified using ISO PHT-S spectra\\
 Sc         & Sc            &  S98, PAH modified using ISO PHT-S spectra\\
 Sd         & Sd            &  S98, PAH modified using ISO PHT-S spectra\\
 M100       & Sbc + HII     &  S98, PAH modified using ISO PHT-S spectra\\
 NGC4027    & Sdm           &  Polletta {\it et al.} (in preparation)\\
 NGC4194    & Im + HII      &  Polletta {\it et al.} (in preparation)\\
 NGC6090(SB)& Starburst     &  S98, PAH modified using ISO PHT-S spectra\\
 Sey1       & Seyfert 1     &  Polletta {\it et al.} (in preparation)\\
 Sey1.5     & Seyfert 1.5   &  Polletta {\it et al.} (in preparation)\\
 Sey2       & Seyfert 2     &  Polletta {\it et al.} (in preparation)\\
 Mrk231     & Type-1 AGN+ULIRG & Berta {\it et al.} (2004)\\
 IRAS19254s & Type-2 AGN+ULIRG & Berta {\it et al.} (2004)\\
 QSO1       & Type-1 QSO    & Composite quasar spectra from the Large\\
            &               & Bright Quasar Survey (Brotherton {\it et
al.} 2001)\\
 LBQS2      & Type-1 QSO    & Composite quasar spectra from the Large\\
            &               & Bright Quasar Survey (Brotherton {\it et
al.} 2001)\\
 QSOIR      & Type-1 QSO    & Same as QSO1 in the optical, IR based on
average\\
            &               & SED of SWIRE type-1 AGNs (Polletta {\it et
al.} in prep)\\
 RLFBQS     & Type-1 QSO    & Composite quasar spectra from the FIRST\\
            &               & Bright Quasar Survey (Brotherton {\it et
al.} 2001)\\
 QSO2       & Type-2 QSO    & FIRST J013435.7-093102 in the optical\\
            &               & (Gregg {\it et al.} 2002)\\
\enddata
\label{tab:types}
\end{deluxetable}
\clearpage

\begin{deluxetable}{lccccccccccccc}
\tabletypesize{\scriptsize}
\tablewidth{0pt} 
\rotate 
\tablecaption{Relevant data on the Spitzer and Chandra Source Sample}
\tablehead{                              
\colhead{id}  &  \colhead{RA} & \colhead{Dec} & \colhead{$S_{3.6}$} & \colhead{$S_{4.5}$} & \colhead{$S_{5.6}$} & \colhead{$S_{8}$} & \colhead{$S_{24}$} & \colhead{$d$} & \colhead{$S_X$} & \colhead{$\delta S_X$} &  \colhead{ty} & \colhead{HR} & \colhead{$L_{0.5-8}$}  \\
\colhead{1}   &  \colhead{2}  & \colhead{3}   & \colhead{4}         & \colhead{5        } & \colhead{6        } & \colhead{7      } & \colhead{8       } & \colhead{9  } & \colhead{10   } & \colhead{11          } &  \colhead{13} & \colhead{14} & \colhead{15         }  \\
}
\startdata
1   & 242.84070 & 54.56754 & 4.90$\pm$0.799 & 15.3$\pm$1.15 & - -           &  -   -        & - -           & 2.94 & 1.24e-14 & 1.53e-15  & 1 & -0.31$\pm$0.10     &   1.45e+44\\
2   & 242.80450 & 54.63007 & 7.94$\pm$0.731 & 16.9$\pm$1.16 & - -           &  -   -        & 120.$\pm$23.1 & 0.69 & 4.24e-15 & 8.81e-16  & 1 & -0.44$\pm$0.16     &   4.95e+43\\
3   & 242.76810 & 54.51849 & 17.4$\pm$0.949 & 14.1$\pm$1.14 & - -           &  -   -        &  -   -        & 0.68 & 4.33e-15 & 8.39e-16  & 3 &  0.03$\pm$0.16     &   2.09e+43\\
4   & 242.74789 & 54.55882 & 66.8$\pm$1.27  & 68.6$\pm$1.5  & 134.$\pm$6.42 & 235.$\pm$6.43 & 637.$\pm$24.9 & 1.15 & 1.61e-14 & 1.50e-15  & 2 & -0.34$\pm$0.08     &   7.77e+43\\
5   & 242.74207 & 54.61105 & 10.7$\pm$0.853 & 9.74$\pm$1.06 & - -           &  -   -        & 45.0$\pm$20.3 & 1.94 & 8.10e-15 & 1.07e-15  & 3 & -0.40$\pm$0.12     &   5.95e+43\\
6   & 242.73192 & 54.64997 & 80.3$\pm$1.23  & 79.2$\pm$1.55 & 110.$\pm$5.21 & 140.$\pm$6.01 & 351.$\pm$19.7 & 2.26 & 2.63e-14 & 1.94e-15  & 2 & -0.25$\pm$0.07     &   6.93e+43\\
8   & 242.73097 & 54.59314 & 52.7$\pm$1.17  & 51.5$\pm$1.41 & 72.6$\pm$5.86 & 109.$\pm$6.2  & 578.$\pm$20.8 & 1.70 & 1.71e-14 & 1.56e-15  & 1 & -0.53$\pm$0.07     &   1.09e+44\\
9   & 242.72926 & 54.53955 & 16.5$\pm$1.06  & - -           & - -           & 31.7$\pm$5.68 & 49.2$\pm$18.4 & 1.62 & 7.41e-15 & 1.03e-15  & 1 & -0.40$\pm$0.12     &   5.86e+43\\
10  & 242.71800 & 54.49846 & 32.6$\pm$0.954 & 28.3$\pm$1.1  & 40.9$\pm$4.56 & 51.5$\pm$4.43 & 167.$\pm$22.5 & 1.38 & 1.66e-15 & 5.29e-16  & 1 & -0.27$\pm$0.28     &   6.36e+42\\
12  & 242.71112 & 54.49831 & 8.82$\pm$0.717 & 12.9$\pm$0.922& - -           &  -   -        &  -   -        & 1.59 & 4.08e-15 & 8.08e-16  & 3 &  0.39$\pm$0.17     &   3.15e+43\\
13  & 242.71115 & 54.66571 & 38.0$\pm$0.858 & 46.6$\pm$1.12 & 52.0$\pm$4.59 & 85.2$\pm$4.55 & 223.$\pm$19.4 & 4.57 & 1.54e-14 & 1.50e-15  & 1 & -0.24$\pm$0.08     &   6.16e+43\\
14  & 242.70901 & 54.50658 & 41.6$\pm$0.867 & 57.7$\pm$1.2  & 58.9$\pm$4.3  & 89.0$\pm$4.87 &  -   -        & 1.42 & 1.95e-14 & 1.83e-15  & 1 & -0.35$\pm$0.08     &   1.60e+44\\
15  & 242.70256 & 54.59798 & 56.3$\pm$1.2   & 64.5$\pm$1.48 & 86.1$\pm$6.08 & 116.$\pm$5.98 & 300.$\pm$23.3 & 0.81 & 1.06e-14 & 1.20e-15  & 1 & -0.62$\pm$0.09     &   1.14e+44\\
16  & 242.69814 & 54.47008 & 39.8$\pm$0.977 & 25.6$\pm$1.23 & 45.5$\pm$5.64 &  -   -        & -  -  	    & 3.23 & 1.47e-15 & 5.09e-16  & 3 &  0.60$\pm$0.21     &   5.65e+42\\
18  & 242.69720 & 54.52565 & 24.9$\pm$0.989 & 14.4$\pm$1.14 & - -           &  -   -        & -  -  	    & 3.07 & 2.43e-15 & 6.23e-16  & 3 & -0.23$\pm$0.22     &   9.30e+42\\
19  & 242.69609 & 54.61655 & 24.9$\pm$0.803 & 14.5$\pm$1.48 & - -           & 56.7$\pm$5.76 & -  -  	    & 1.26 & 6.55e-15 & 9.72e-16  & 1 & -0.46$\pm$0.13     &   2.78e+44\\
20  & 242.69415 & 54.59382 & 552.$\pm$3.31  & 355.$\pm$3.42 & 470.$\pm$12.3 & 2597$\pm$13   & 1919$\pm$30.9 & 1.21 & 2.31e-15 & 6.11e-16  & 3 & -0.33$\pm$0.24     &   2.35e+40\\
21  & 242.69202 & 54.39124 & 137.$\pm$2.02  & 105.$\pm$2.3  & 55.0$\pm$5.75 & 93.4$\pm$5.86 & 217.$\pm$21.4 & 1.53 & 1.86e-14 & 1.85e-15  & 3 & -0.43$\pm$0.08     &   7.12e+43\\
22  & 242.68829 & 54.49787 & 50.1$\pm$1.22  & 42.5$\pm$1.39 & - -           &  -   -        & 80.3$\pm$22   & 0.37 & 2.40e-15 & 6.18e-16  & 3 &  0.92$\pm$0.19     &   4.66e+42\\
23  & 242.68791 & 54.60337 & 479.$\pm$2.38  & 507.$\pm$2.76 & 640.$\pm$7.43 & 667.$\pm$6.48 & 1357$\pm$23.2 & 1.49 & 3.59e-13 & 6.78e-15  & 1 & -0.43$\pm$0.01     &   4.02e+44\\
25  & 242.68373 & 54.60024 & 26.1$\pm$0.838 & 25.5$\pm$1.21 & 56.2$\pm$4.73 &  -   -        & 114.$\pm$22.1 & 1.56 & 1.88e-14 & 1.57e-15  & 2 & -0.30$\pm$0.10     &   4.85e+44\\
26  & 242.67850 & 54.45293 & 106.$\pm$1.64  & 74.0$\pm$1.71 & 60.6$\pm$5.83 & 86.8$\pm$5.83 & 400.$\pm$22.9 & 0.79 & 3.31e-15 & 7.38e-16  & 2 & -0.54$\pm$0.20     &   6.07e+42\\
27  & 242.67313 & 54.49701 & 18.3$\pm$1.05  & - -           & - -           & 79.5$\pm$5.8  & 346.$\pm$19.9 & 1.88 & 3.03e-15 & 7.47e-16  & 1 & -0.33$\pm$0.23     &   5.87e+43\\
28  & 242.67230 & 54.57432 & 17.1$\pm$0.889 & 18.1$\pm$1.3  & - -           &  -   -        & 188.$\pm$22.9 & 1.03 & 1.09e-15 & 4.15e-16  & 1 & -1.03$\pm$0.36     &   9.98e+42\\
30  & 242.66684 & 54.66679 & 39.2$\pm$1.09  & 18.9$\pm$1.28 & - -           &  -   -        & 129.$\pm$19.9 & 2.67 & 9.45e-15 & 1.17e-15  & 3 & -0.28$\pm$0.11     &   5.47e+42\\
31  & 242.66272 & 54.62717 & 73.7$\pm$1.35  & 93.7$\pm$1.67 & 252.$\pm$10.2 & 295.$\pm$6.38 & 1105$\pm$22   & 2.0  & 7.81e-15 & 1.04e-15  & 2 & -0.44$\pm$0.12     &   6.56e+43\\
34  & 242.64740 & 54.54727 & 30.2$\pm$0.999 & 23.7$\pm$1.31 & 31.9$\pm$5.66 &  -   -        & 9.85$\pm$21   & 0.74 & 4.95e-15 & 8.16e-16  & 3 & -0.48$\pm$0.15     &   3.62e+43\\
38  & 242.63971 & 54.52485 & 85.2$\pm$1.39  & 68.9$\pm$1.65 & 35.6$\pm$5.53 & 72.0$\pm$5.79 & 317.$\pm$21   & 3.53 & 1.86e-15 & 5.08e-16  & 2 &  0.54$\pm$0.26     &   6.92e+42\\
39  & 242.63295 & 54.53443 & -99.$\pm$-     & 36.2$\pm$1.35 & 40.0$\pm$5.68 &  -   -        & - -           & 1.71 & 7.65e-15 & 1.13e-15  & 3 & -0.09$\pm$0.14     &   8.75e+43\\
40  & 242.62520 & 54.52821 & 19.5$\pm$0.987 & 21.8$\pm$1.37 & 74.8$\pm$5.92 & 83.8$\pm$5.84 & 289.$\pm$21.3 & 2.00 & 1.57e-14 & 1.41e-15  & 1 & -0.32$\pm$0.08     &   2.91e+43\\
41  & 242.61415 & 54.50594 & 32.9$\pm$0.655 & 24.8$\pm$1.25 & 40.6$\pm$4.05 &  -   -        & - -           & 4.99 & 1.21e-15 & 4.47e-16  & 3 &  0.87$\pm$0.30     &   4.65e+42\\
45  & 242.59665 & 54.50228 & 38.0$\pm$1.1   & 29.0$\pm$1.36 & 27.9$\pm$5.63 & 73.6$\pm$5.79 & 322.$\pm$19.6 & 1.65 & 5.90e-15 & 9.21e-16  & 2 &  0.46$\pm$0.17     &   7.91e+42\\
46  & 242.59344 & 54.53010 & 27.5$\pm$1.03  & 26.8$\pm$1.3  & - -           &  -   -        & - -           & 1.06 & 1.48e-15 & 4.80e-16  & 2 &  0.39$\pm$0.27     &   7.42e+42\\
48  & 242.59050 & 54.51771 & 87.5$\pm$1.16  & 114.$\pm$1.82 & 140.$\pm$6.02 & 228.$\pm$6.36 & 498.$\pm$24.3 & 1.48 & 3.17e-14 & 1.98e-15  & 1 & -0.43$\pm$0.05     &   6.09e+44\\
50  & 242.58466 & 54.50531 & 111.$\pm$1.56  & 143.$\pm$1.94 & 162.$\pm$7.15 & 267.$\pm$6.32 & 861.$\pm$24   & 0.97 & 5.35e-14 & 2.59e-15  & 1 & -0.52$\pm$0.04     &   8.64e+43\\
51  & 242.58406 & 54.49353 & 14.1$\pm$1.19  & 19.3$\pm$1.23 & - -           &  -   -        & - -           & 1.99 & 1.86e-15 & 5.52e-16  & 3 &  0.02$\pm$0.25     &   1.87e+43\\
53  & 242.57802 & 54.54125 & 4.92$\pm$0.786 & 8.33$\pm$1.09 & - -           &  -   -        & 14.4$\pm$4.73 & 2.38 & 1.28e-15 & 4.47e-16  & 2 &  0.51$\pm$0.28     &   1.85e+43\\
54  & 242.56976 & 54.52677 & 8.98$\pm$0.571 & 10.6$\pm$0.87 & - -           &  -   -        & 219.$\pm$19.7 & 1.02 & 1.32e-15 & 4.50e-16  & 2 &  0.34$\pm$0.25     &   1.44e+43\\
55  & 242.56297 & 54.59608 & 60.8$\pm$1.14  & 39.2$\pm$1.2  & - -           &  -   -        & -  -  	    & 0.57 & 3.34e-15 & 6.73e-16  & 3 & -0.38$\pm$0.18     &   1.92e+43\\
56  & 242.56113 & 54.46746 & 12.1$\pm$0.737 & 15.5$\pm$0.948& - -           &  -   -        & -  -  	    & 1.03 & 3.24e-15 & 7.10e-16  & 3 &  0.03$\pm$0.23     &   2.63e+43\\
57  & 242.56024 & 54.63151 & 37.0$\pm$0.978 & 40.7$\pm$1.22 & 57.4$\pm$4.61 & 99.4$\pm$5.27 & 81.5$\pm$22.2 & 2.92 & 2.79e-15 & 6.65e-16  & 2 & -0.12$\pm$0.22     &   9.97e+43\\
58  & 242.55421 & 54.58305 & 15.4$\pm$0.738 & 20.7$\pm$1.2  & 46.6$\pm$4.56 &  -   -        & - -           & 1.73 & 1.08e-15 & 4.16e-16  & 1 &  0.45$\pm$0.33     &   3.20e+42\\
59  & 242.55330 & 54.46565 & 49.0$\pm$0.966 & 59.9$\pm$1.4  & 90.$\pm$5.33  & 118.$\pm$5.41 & 666.$\pm$27.3 & 0.35 & 2.83e-14 & 1.95e-15  & 1 & -0.38$\pm$0.06     &   4.41e+43\\
61  & 242.53992 & 54.54564 & 51.5$\pm$0.964 & 26.5$\pm$1.26 & - -           &  -   -        & - -           & 1.35 & 3.21e-15 & 6.78e-16  & 3 & -0.44$\pm$0.20     &   4.95e+42\\
62  & 242.53745 & 54.56409 & 80.2$\pm$1.21  & 77.1$\pm$1.27 & 76.2$\pm$4.97 & 82.2$\pm$4.75 & 341.$\pm$19.5 & 1.51 & 1.11e-14 & 1.80e-15  & 2 & -0.67$\pm$0.14     &   2.08e+44\\
64  & 242.53362 & 54.55189 & 33.4$\pm$0.954 & 38.8$\pm$1.11 & 55.5$\pm$5.18 & 57.9$\pm$4.35 & 471.$\pm$22.7 & 1.60 & 6.30e-15 & 9.03e-16  & 1 & -0.89$\pm$0.12     &   3.50e+44\\
65  & 242.53102 & 54.50165 & 6.42$\pm$0.52  & 9.69$\pm$0.783& - -           &  -   -        & - -           & 0.91 & 4.78e-15 & 1.01e-15  & 1 &  0.09$\pm$0.20     &   9.12e+43\\
66  & 242.52936 & 54.62281 & 4.18$\pm$0.66  & 7.97$\pm$1.12 & - -           &  -   -        & 130.$\pm$20.5 & 2.90 & 1.88e-15 & 5.42e-16  & 1 & -0.87$\pm$0.25     &   1.05e+44\\
67  & 242.52798 & 54.54510 & 55.6$\pm$1.07  & 39.3$\pm$1.35 & 49.6$\pm$4.87 &  -   -        & 248.$\pm$24.4 & 1.64 & 2.22e-14 & 1.67e-15  & 1 & -0.44$\pm$0.06     &   7.25e+43\\
68  & 242.52010 & 54.58688 & 10.7$\pm$0.841 & 11.8$\pm$1.11 & - -           &  -   -        & - -           & 1.56 & 1.99e-15 & 5.59e-16  & 1 & -0.59$\pm$0.25     &   1.61e+43\\
69  & 242.51291 & 54.60775 & 243.$\pm$1.5   & 221.$\pm$1.84 & 168.$\pm$4.91 & 581.$\pm$7.21 & 1255$\pm$21.1 & 2.12 & 5.95e-14 & 2.73e-15  & 3 & -0.40$\pm$0.04     &   1.25e+43\\
71  & 242.50871 & 54.42397 & 66.3$\pm$1.14  & 56.9$\pm$1.38 & 79.4$\pm$5.1  & 81.2$\pm$4.91 & 301.$\pm$20.5 & 1.84 & 1.58e-14 & 1.64e-15  & 1 & -0.44$\pm$0.09     &   2.42e+43\\
72  & 242.50519 & 54.63092 & 10.8$\pm$0.833 & 7.96$\pm$1.09 & - -           &  -   -        & 90.7$\pm$22   & 1.13 & 2.03e-15 & 5.63e-16  & 3 &  0.70$\pm$0.22     &   7.81e+42\\
73  & 242.50346 & 54.65526 & 49.7$\pm$0.954 & 42.2$\pm$1.2  & 76.1$\pm$4.84 & 108.$\pm$4.83 & 438.$\pm$22.2 & 0.84 & 3.16e-15 & 7.37e-16  & 2 &  0.24$\pm$0.21     &   1.07e+43\\
75  & 242.49825 & 54.55404 & 64.2$\pm$1.28  & 103.$\pm$1.68 & 133.$\pm$6.9  & 210.$\pm$6.37 & 613.$\pm$21.4 & 1.83 & 1.61e-14 & 1.47e-15  & 1 & -0.46$\pm$0.08     &   1.99e+44\\
76  & 242.49693 & 54.46495 & 15.9$\pm$0.751 & - -           & - -           &  -   -        & 257.$\pm$18.2 & 4.76 & 4.90e-15 & 1.31e-15  & 1 & -0.27$\pm$0.27     &   4.95e+43\\
77  & 242.48674 & 54.57907 & 7.06$\pm$0.653 & 10.2$\pm$0.811& - -           &  -   -        & - -           & 2.68 & 3.37e-15 & 9.42e-16  & 1 & -0.97$\pm$0.29     &   1.68e+44\\
78  & 242.48642 & 54.58611 & 42.3$\pm$1.04  & 53.3$\pm$1    & 73.3$\pm$5.22 & 108.$\pm$4.17 & 399.$\pm$24.1 & 1.38 & 3.79e-15 & 7.80e-16  & 2 & -0.45$\pm$0.18     &   6.91e+43\\
81  & 242.46735 & 54.59394 & 14.2$\pm$0.986 & 25.5$\pm$1.19 & 35.8$\pm$5.61 & 50.8$\pm$5.13 & 162.$\pm$23.7 & 1.82 & 5.75e-15 & 8.94e-16  & 1 & -0.61$\pm$0.13     &   1.48e+44\\
82  & 242.46300 & 54.46716 & 33.3$\pm$1.06  & 25.8$\pm$1.36 & 28.9$\pm$5.49 &  -   -        & - -           & 1.79 & 9.49e-15 & 1.19e-15  & 2 & -0.49$\pm$0.11     &   5.84e+43\\
83  & 242.46237 & 54.60486 & 14.8$\pm$0.865 & 14.0$\pm$1.2  & - -           & 75.8$\pm$5.97 & 288.$\pm$22.5 & 1.89 & 4.83e-15 & 8.92e-16  & 1 &  0.25$\pm$0.17     &   1.06e+43\\
84  & 242.45204 & 54.71840 & 76.4$\pm$1.33  & 103.$\pm$1.69 & 133.$\pm$6.46 & 202.$\pm$6.46 & 492.$\pm$22.4 & 4.94 & 4.29e-14 & 2.63e-15  & 1 & -0.36$\pm$0.05     &   6.71e+43\\
86  & 242.43242 & 54.63044 & 156.$\pm$1.81  & 111.$\pm$2.03 & 73.8$\pm$5.9  & 80.0$\pm$5.8  & 8.31$\pm$22.7 & 2.45 & 3.75e-15 & 7.81e-16  & 3 & -0.06$\pm$0.17     &   7.89e+42\\
87  & 242.42087 & 54.67013 & 42.0$\pm$0.873 & 56.7$\pm$1.21 & 56.2$\pm$5.24 & 88.3$\pm$4.77 & 224.$\pm$20.3 & 1.98 & 1.98e-14 & 1.75e-15  & 1 & -0.46$\pm$0.08     &   7.95e+43\\
89  & 242.40501 & 54.67556 & 89.6$\pm$1.5   & 50.9$\pm$1.24 & 58.4$\pm$5.15 & 83.4$\pm$5.26 & 219.$\pm$20.8 & 0.61 & 4.31e-14 & 2.57e-15  & 2 & -0.28$\pm$0.05     &   1.49e+44\\
90  & 242.40094 & 54.63650 & 43.9$\pm$1.11  & 50.0$\pm$1.39 & - -           & 78.4$\pm$6.22 & 325.$\pm$20.9 & 1.17 & 1.07e-14 & 1.27e-15  & 1 & -0.46$\pm$0.10     &   7.34e+43\\
92  & 242.39139 & 54.61434 & 185.$\pm$1.91  & 177.$\pm$2.08 & 215.$\pm$8.76 & 314.$\pm$6.39 & 889.$\pm$22.5 & 1.21 & 6.23e-15 & 9.73e-16  & 2 &  0.42$\pm$0.12     &   2.23e+43\\
93  & 242.38696 & 54.53603 & 45.7$\pm$0.989 & 32.8$\pm$1.51 & - -           & 55.2$\pm$5.74 & 71.3$\pm$23.6 & 1.21 & 1.57e-14 & 1.50e-15  & 2 & -0.33$\pm$0.09     &   1.43e+44\\
94  & 242.38441 & 54.53204 & 11.1$\pm$0.692 & 10.1$\pm$0.906& - -           &  -   -        & 197.$\pm$22.7 & 1.42 & 6.00e-15 & 9.81e-16  & 1 & -0.27$\pm$0.15     &   1.26e+44\\
95  & 242.31757 & 54.53606 & 53.1$\pm$1.2   & 30.6$\pm$1.13 & - -           &  -   -        & - -           & 3.22 & 6.28e-15 & 1.06e-15  & 3 & -0.32$\pm$0.13     &   1.49e+43\\
96  & 242.83466 & 54.58537 & 956.$\pm$3.36  & 630.$\pm$3.49 & 428.$\pm$10.3 & 378.$\pm$9.3  & 474.$\pm$26   & 1.2  & 4.53e-15 & 8.99e-16  & 4 &  1.11$\pm$0.09     &   3.99e+41\\
97  & 242.78481 & 54.54735 & 40.3$\pm$0.978 & 31.8$\pm$1.14 & -  -  	    &  -   -        & 448.$\pm$23.1 & 0.60 & 3.93e-15 & 7.91e-16  & 3 & -0.64$\pm$0.18     &   1.93e+43\\
98  & 242.77934 & 54.59368 & 8.09$\pm$0.855 & 12.9$\pm$1.12 & -  -  	    &  -   -        & -  -  	    & 1.77 & 2.18e-15 & 6.03e-16  & 3 & -0.13$\pm$0.22     &   4.05e+43\\
99  & 242.76178 & 54.49972 & 9.75$\pm$0.832 & 7.80$\pm$1.09 & -  -  	    &  -   -        & -  -  	    & 0.58 & 6.99e-15 & 1.16e-15  & 1 & -0.15$\pm$0.15     &   3.10e+43\\
100 & 242.75789 & 54.64089 & 49.9$\pm$0.863 & 34.8$\pm$0.973& -  -  	    & 18.1$\pm$4.05 & -  -  	    & 3.31 & 3.17e-15 & 7.33e-16  & 3 & -0.11$\pm$0.18     &   1.28e+43\\
104 & 242.69458 & 54.41047 & 40.3$\pm$0.891 & 34.5$\pm$1.18 & -  -  	    & 64.8$\pm$5.15 & -  -  	    & 1.69 & 1.14e-14 & 1.46e-15  & 2 & -0.39$\pm$0.10     &   4.41e+43\\
106 & 242.57269 & 54.53015 & 9.10$\pm$0.599 & 13.9$\pm$0.802& 21.0$\pm$4.16 &  -   -        & 198.$\pm$19.7 & 1.31 & 1.38e-15 & 4.64e-16  & 1 &  0.20$\pm$0.29     &   1.89e+43\\
108 & 242.52039 & 54.44347 & 18.7$\pm$0.797 & 20.5$\pm$1.12 & 22.1$\pm$4.21 &  -   -        & - -           & 2.55 & 6.38e-15 & 1.07e-15  & 2 & -0.42$\pm$0.16     &   2.80e+43\\
112 & 242.45084 & 54.60307 & 120.$\pm$1.53  & 89.4$\pm$1.6  & 80.4$\pm$5.99 & 153.$\pm$6.25 & 1303$\pm$23.4 & 0.21 & 2.41e-15 & 6.47e-16  & 2 &  0.80$\pm$0.20     &   4.99e+42\\
113 & 242.85968 & 54.59127 & 9.67$\pm$0.755 & 17.3$\pm$0.822& 40.1$\pm$4.59 & 71.0$\pm$3.72 & 62.9$\pm$23   & 0.86 & 5.14e-15 & 1.09e-15  & 1 & -0.15$\pm$0.16     &   4.00e+43\\
114 & 242.76576 & 54.55086 & 120.$\pm$1.78  & 81.1$\pm$2    & 71.0$\pm$5.82 &  -   -        & - -           & 1.30 & 2.11e-15 & 6.23e-16  & 4 &  0.05$\pm$0.21     &   2.16e+42\\
115 & 242.62517 & 54.40055 & 115.$\pm$1.37  & 91.5$\pm$1.57 & 61.3$\pm$5.77 & 54.0$\pm$4.71 & 169.$\pm$23.7 & 4.56 & 3.84e-15 & 9.22e-16  & 4 & -0.55$\pm$0.18     &   2.46e+40\\
116 & 242.59047 & 54.43574 & 58.8$\pm$1.15  & 26.0$\pm$1.39 & -  -  	    &  -   -        & 158.$\pm$23.7 & 1.30 & 1.65e-15 & 5.62e-16  & 3 &  0.79$\pm$0.18     &   4.26e+42\\
117 & 242.57739 & 54.45892 & 32.6$\pm$1.06  & 31.8$\pm$1.36 & -  -  	    &  -   -        & - -           & 2.45 & 2.42e-15 & 6.39e-16  & 3 &  0.20$\pm$0.21     &   2.12e+43\\
118 & 242.52217 & 54.65250 & 51.1$\pm$1.06  & 31.8$\pm$1.42 & -  -  	    &  -   -        & 118.$\pm$23.1 & 2.35 & 1.73e-15 & 5.44e-16  & 2 & -0.22$\pm$0.26     &   1.25e+43\\
119 & 242.47356 & 54.63194 & 117.$\pm$1.43  & 64.9$\pm$1.13 & 65.1$\pm$4.78 &  -   -        & 108.$\pm$23.2 & 3.30 & 1.74e-15 & 5.47e-16  & 3 &  0.72$\pm$0.18     &   3.26e+42\\
121 & 242.38539 & 54.57699 & 25.5$\pm$0.801 & 18.7$\pm$0.856& - -           &  -   -        & - -           & 3.65 & 1.83e-15 & 5.76e-16  & 3 &  0.79$\pm$0.14     &   6.10e+42\\
124 & 242.88104 & 54.61811 & 69.1$\pm$1.15  & 64.8$\pm$1.47 & 86.6$\pm$4.92 & 95.1$\pm$5.83 & 164.$\pm$23.6 & 1.66 & 2.89e-15 & 8.90e-16  & 2 & -0.38$\pm$0.18     &   9.99e+41\\
127 & 242.75615 & 54.50836 & - -            & 21.2$\pm$1.21 & -  -  	    &  -   -        & 226.$\pm$22.1 & 4.17 & 2.10e-15 & 6.83e-16  & 1 & -0.54$\pm$0.24     &   4.64e+42\\
128 & 242.68021 & 54.56771 & 14.8$\pm$0.855 & 15.3$\pm$1.18 & -  -  	    &  -   -        & 110.$\pm$19   & 4.8  & 1.61e-15 & 5.20e-16  & 3 & -0.37$\pm$0.26     &   1.11e+43\\
131 & 242.43211 & 54.64696 & 15.2$\pm$0.959 & 18.6$\pm$1.17 & -  -  	    &  -   -        & - -           & 2.10 & 1.86e-15 & 5.63e-16  & 4 & -0.04$\pm$0.18     &   2.74e+43\\
132 & 242.33562 & 54.51714 & 62.7$\pm$1.26  & 41.7$\pm$1.38 & 52.1$\pm$5.69 & 80.3$\pm$5.87 & 103.$\pm$23.5 & 2.29 & 2.32e-15 & 6.70e-16  & 2 &  0.00$\pm$0.17     &   4.43e+42\\
133 & 242.32001 & 54.46970 & 17.1$\pm$0.948 & 13.3$\pm$1.18 & - -           &  -   -        & 111.$\pm$22.3 & 4.47 & 5.05e-15 & 1.07e-15  & 3 &  0.25$\pm$0.13     &   2.35e+43\\
136 & 242.53204 & 54.64293 & 237.$\pm$2.3   & 145.$\pm$2.51 & 98.6$\pm$5.96 &  -   -        & - -           & 1.31 & 2.17e-15 & 6.20e-16  & 4 & -0.22$\pm$0.24     &   3.49e+40\\
139 & 242.42795 & 54.45282 & 20.6$\pm$0.815 & 21.5$\pm$1.05 & 64.1$\pm$5.19 & 83.2$\pm$4.89 & 337.$\pm$22.3 & 3.89 & 2.41e-15 & 6.81e-16  & 1 & -0.34$\pm$0.19     &   4.89e+42\\
140 & 242.74313 & 54.64743 & 10.6$\pm$0.713 & 7.14$\pm$1.06 & - -           &  -   -        & 72.0$\pm$24.3 & 3.13 & 1.90e-15 & 5.75e-16  & 3 & -0.66$\pm$0.28     &   2.96e+42\\
141 & 242.42952 & 54.69754 & 37.9$\pm$1.05  & 52.7$\pm$1.46 & 90.2$\pm$5.38 & 71.8$\pm$5.85 & 442.$\pm$22.1 & 3.00 & 2.55e-15 & 7.72e-16  & 1 & -1.10$\pm$0.20     &   4.29e+42\\
142 & 242.80632 & 54.60314 & 24.0$\pm$0.978 & 28.2$\pm$1.31 & -  -  	    &  -   -        & 435.$\pm$20.2 & 1.29 & 1.97e-15 & 6.44e-16  & 1 & -0.96$\pm$0.25     &   2.53e+44\\
143 & 242.41783 & 54.62019 & 67.5$\pm$1.28  & 46.5$\pm$1.21 & -  -  	    &  -   -        & 99.3$\pm$22.8 & 3.44 & 1.52e-15 & 5.27e-16  & 2 &  1.21$\pm$0.24     &   2.55e+42\\
\enddata
\tablenotetext{~}{Col. 1: id number as in Manners et al. (2003); Col. 2 \& 3: RA and Dec; 
Col. 4: IRAC flux at 3.6 $\mu$m; Col. 5: IRAC flux at 4.5 $\mu$m; Col. 6: 
IRAC flux at 5.6 $\mu$m; Col. 7: IRAC flux at 8 $\mu$m; Col. 8: MIPS flux at 24 $\mu$m;
Col. 9: angular distance of the Chandra source to the Spitzer identification;
Col. 10 \& 11: Chandra 0.5 to 8 keV flux (erg/cm$^2$/s) and error; Col. 13: spectral type (1: type-1 AGN, 2: type-2 AGN, 
3: galaxy, 4: early-type galaxy) based on the optical/IR SED; Col. 14: X-ray 
hardness ratio; Col. 15: 0.5-8 keV luminosity.
}
\label{seds}
\end{deluxetable}						
\clearpage								
										
\end{document}